\input harvmac
\input epsf

\newcount\figno
\figno=0
\def\fig#1#2#3{
\par\begingroup\parindent=0pt\leftskip=1cm\rightskip=1cm\parindent=0pt
\baselineskip=12pt
\global\advance\figno by 1
\midinsert
\epsfxsize=#3
\centerline{\epsfbox{#2}}
\vskip 14pt

{\bf Fig. \the\figno:} #1\par
\endinsert\endgroup\par
}
\def\figlabel#1{\xdef#1{\the\figno}}
\def\encadremath#1{\vbox{\hrule\hbox{\vrule\kern8pt\vbox{\kern8pt
\hbox{$\displaystyle #1$}\kern8pt}
\kern8pt\vrule}\hrule}}

\overfullrule=0pt

\noblackbox
\parskip=1.5mm

\def\Title#1#2{\rightline{#1}\ifx\answ\bigans\nopagenumbers\pageno0
\else\pageno1\vskip.5in\fi \centerline{\titlefont #2}\vskip .3in}

\font\caps=cmcsc10

\noblackbox
\parskip=1.5mm



           \def\CO{{\cal O}} 
\def\CA{{\cal A}}   
\def\CL{{\cal L}}   
 \def\CR{{\cal R}}


\def\dj{\hbox{d\kern-0.347em \vrule width 0.3em height 1.252ex depth
-1.21ex \kern 0.051em}}

\def\half{{1\over 2}\,}

\def\ket{\rangle}
\def\bra{\langle}

\def\tX{\widetilde X}
\def\tphi{\widetilde \phi}
\def\tpi{\widetilde \pi}

\def\tPsi{\widetilde \Psi}
\def\tomega{\widetilde \omega}

\def\pt{\partial}

\def\Dirac{\,\raise.15ex\hbox{/}\mkern-13.5mu D}
\def\dirac{\,\raise.15ex\hbox{/}\kern-.57em \partial}
\def\aslash{\,\raise.15ex\hbox{/}\mkern-13.5mu A}
\def\btX{{\bf {\widetilde X}}}

\def\shalf{{\ifinner {\textstyle {1 \over 2}}\else {1 \over 2} \fi}}
\def\sshalf{{\ifinner {\scriptstyle {1 \over 2}}\else {1 \over 2} \fi}}
\def\sfourth{{\ifinner {\textstyle {1 \over 4}}\else {1 \over 4} \fi}}
\def\sthreehalfs{{\ifinner {\textstyle {3 \over 2}}\else {3 \over 2} \fi}}
\def\sdhalfs{{\ifinner {\textstyle {d \over 2}}\else {d \over 2} \fi}}
\def\sdmtwohalfs{{\ifinner {\textstyle {d-2 \over 2}}\else {d-2 \over 2} \fi}}
\def\sdmasonehalfs{{\ifinner {\textstyle {d+1 \over 2}}\else {d+1 \over 2} \fi}}
\def\sdmasthreehalfs{{\ifinner {\textstyle {d+3 \over 2}}\else {d+3 \over 2} \fi}}
\def\sdmastwohalfs{{\ifinner {\textstyle {d+2 \over 2}}\else {d+2 \over 2} \fi}}


\lref\rt{
  S.~Ryu and T.~Takayanagi, ``Holographic derivation of entanglement entropy from AdS/CFT,'' Phys.\ Rev.\ Lett.\ {\bf
    96}, 181602 (2006) [arXiv:hep-th/0603001].

  ``Aspects of holographic entanglement entropy,'' JHEP {\bf 0608}, 045 (2006) [arXiv:hep-th/0605073].}

 \lref\eternalmalda{
 J.~M.~Maldacena,
  ``Eternal black holes in anti-de Sitter,''
  JHEP {\bf 0304}, 021 (2003)
  [hep-th/0106112].
  }
  
  \lref\maldahartman{
  T.~Hartman and J.~Maldacena,
  ``Time Evolution of Entanglement Entropy from Black Hole Interiors,''
  JHEP {\bf 1305}, 014 (2013)
  [arXiv:1303.1080 [hep-th]].
  }
  
  \lref\epr{
   J.~Maldacena and L.~Susskind,
  ``Cool horizons for entangled black holes,''
  Fortsch.\ Phys.\  {\bf 61}, 781 (2013)
  [arXiv:1306.0533 [hep-th]].
  }

  \lref\mvr{
   M.~Van Raamsdonk,
  ``Building up spacetime with quantum entanglement,''
  Gen.\ Rel.\ Grav.\  {\bf 42}, 2323 (2010)
  [Int.\ J.\ Mod.\ Phys.\ D {\bf 19}, 2429 (2010)]
  [arXiv:1005.3035 [hep-th]].
  
  M.~Van Raamsdonk,
  ``A patchwork description of dual spacetimes in AdS/CFT,''
  Class.\ Quant.\ Grav.\  {\bf 28}, 065002 (2011).
  }

  \lref\llast{
   L.~Susskind,
  ``The Typical-State Paradox: Diagnosing Horizons with Complexity,''
  arXiv:1507.02287 [hep-th].
  }
  
  \lref\enen{
  L.~Susskind,
  ``Entanglement is not Enough,''
  arXiv:1411.0690 [hep-th].
  }
  
\lref\stanfordsus{D.~Stanford and L.~Susskind,
  ``Complexity and Shock Wave Geometries,''
  Phys.\ Rev.\ D {\bf 90}, no. 12, 126007 (2014)
  [arXiv:1406.2678 [hep-th]].
}

\lref\bridges{
L.~Susskind and Y.~Zhao,
  ``Switchbacks and the Bridge to Nowhere,''
  arXiv:1408.2823 [hep-th].
  }
\lref\suscomple{
L.~Susskind,
  ``Computational Complexity and Black Hole Horizons,''
  arXiv:1403.5695 [hep-th], arXiv:1402.5674 [hep-th].
  }
 
  \lref\tensors{
 B.~Swingle,
  ``Entanglement Renormalization and Holography,''
  Phys.\ Rev.\ D {\bf 86}, 065007 (2012)
  [arXiv:0905.1317 [cond-mat.str-el]].
  
  G.~Evenbly and G.~Vidal, ``Tensor Network States and Geometry,"
  Journal of Statistical Physics 145 (2011) 891Ð918, [arXiv:1106.1082 [quant-ph]].
  
   B.~Swingle,
   ``Constructing holographic spacetimes using entanglement renormalization,''
  arXiv:1209.3304 [hep-th].
  
  X.~L.~Qi,
  ``Exact holographic mapping and emergent space-time geometry,''
  arXiv:1309.6282 [hep-th].
  
  J.~I.~Latorre and G.~Sierra,
  ``Holographic codes,''
  arXiv:1502.06618 [quant-ph].
  
  F.~Pastawski, B.~Yoshida, D.~Harlow and J.~Preskill,
  ``Holographic quantum error-correcting codes: Toy models for the bulk/boundary correspondence,''
  JHEP {\bf 1506}, 149 (2015)
  [arXiv:1503.06237 [hep-th]].
  }
  
\lref\action{
A.~R.~Brown, D.~A.~Roberts, L.~Susskind, B.~Swingle and Y.~Zhao,
  ``Complexity Equals Action,''
  arXiv:1509.07876 [hep-th].
  }

  \lref\lasttaka{
  M.~Miyaji, T.~Numasawa, N.~Shiba, T.~Takayanagi and K.~Watanabe,
  ``Gravity Dual of Quantum Information Metric,''
  arXiv:1507.07555 [hep-th].
  }
  
  \lref\alisha{
  M.~Alishahiha,
  ``Holographic Complexity,''
  arXiv:1509.06614 [hep-th].
  }

\lref\usu{
  J.~L.~F.~Barb\'on and E.~Rabinovici,
  ``Holography of AdS vacuum bubbles,''
  JHEP {\bf 1004}, 123 (2010)
  [arXiv:1003.4966 [hep-th]].}

 \lref\adscft{
  J.~M.~Maldacena,
  ``The large N limit of superconformal field theories and supergravity,''
  Adv.\ Theor.\ Math.\ Phys.\  {\bf 2}, 231 (1998)
  [Int.\ J.\ Theor.\ Phys.\  {\bf 38}, 1113 (1999)]
  [arXiv:hep-th/9711200].
 S.~S.~Gubser, I.~R.~Klebanov and A.~M.~Polyakov,
  ``Gauge theory correlators from non-critical string theory,''
  Phys.\ Lett.\  B {\bf 428}, 105 (1998)
  [arXiv:hep-th/9802109].
 E.~Witten,
  ``Anti-de Sitter space and holography,''
  Adv.\ Theor.\ Math.\ Phys.\  {\bf 2}, 253 (1998)
  [arXiv:hep-th/9802150].
  }

   \lref\roberto{
   R.~Emparan,
  ``AdS / CFT duals of topological black holes and the entropy of zero energy states,''
  JHEP {\bf 9906}, 036 (1999)
  [hep-th/9906040].
  }

\lref\mukund{
  D.~Marolf, M.~Rangamani and M.~Van Raamsdonk,
  ``Holographic models of de Sitter QFTs,''
  arXiv:1007.3996 [hep-th].}

\lref\insightfull{
  G.~Horowitz, A.~Lawrence and E.~Silverstein,
  ``Insightful D-branes,''
  JHEP {\bf 0907}, 057 (2009)
  [arXiv:0904.3922 [hep-th]].}


\lref\har{
  D.~Harlow,
  ``Metastability in Anti de Sitter Space,''
  arXiv:1003.5909 [hep-th].}

\lref\magan{
  J.~L.~F.~Barb\'on and J.~Mart\'{\i}nez-Mag\'an,
  ``Spontaneous fragmentation of topological black holes,''
  JHEP {\bf 1008}, 031 (2010)
  [arXiv:1005.4439 [hep-th]].
}

\lref\craps{
A.~Bernamonti and B.~Craps,
  ``D-Brane Potentials from Multi-Trace Deformations in AdS/CFT,''
  JHEP {\bf 0908}, 112 (2009)
  [arXiv:0907.0889 [hep-th]].
}

\lref\fuertes{
 J.~L.~F.~Barbon and C.~A.~Fuertes,
  ``A Note on the extensivity of the holographic entanglement entropy,''
  JHEP {\bf 0805}, 053 (2008)
  [arXiv:0801.2153 [hep-th]].
  }

 \lref\rcmt{
  S.~A.~Hartnoll,
  ``Lectures on holographic methods for condensed matter physics,''
  Class.\ Quant.\ Grav.\  {\bf 26}, 224002 (2009)
  [arXiv:0903.3246 [hep-th]].
  }
  
  \lref\marpol{
  D.~Marolf and J.~Polchinski,
  ``Gauge/Gravity Duality and the Black Hole Interior,''
  Phys.\ Rev.\ Lett.\  {\bf 111}, 171301 (2013)
  [arXiv:1307.4706 [hep-th]].
  }
  
  \lref\icec{
  V.~Balasubramanian, M.~Berkooz, S.~F.~Ross and J.~Simon,
  ``Black Holes, Entanglement and Random Matrices,''
  Class.\ Quant.\ Grav.\  {\bf 31}, 185009 (2014)
  [arXiv:1404.6198 [hep-th]].
  }
  
  \lref\noise{
  J.~L.~F.~Barbon and E.~Rabinovici,
  ``Geometry And Quantum Noise,''
  Fortsch.\ Phys.\  {\bf 62}, 626 (2014)
  [arXiv:1404.7085 [hep-th]].
  }
  
  \lref\hypobh{
  J.P.~Lemos,
``Cylindrical black hole in general relativity,"
Phys. Lett. {\bf B353}, 46 (1994)
gr-qc/9404041;

J.P.~Lemos,
``Two-dimensional black holes and planar general relativity,"
Class. Quant. Grav. {\bf 12}, 1081 (1995)
gr-qc/9407024;

S.~Aminneborg, I.~Bengtsson, S.~Holst and P.~Peld{\'a}n,
``Making anti-de Sitter black holes,"
Class. Quant. Grav. {\bf 13}, 2707 (1996)
gr-qc/9604005;

R.B.~Mann,
``Pair production of topological anti-de Sitter black holes,"
Class. Quant. Grav. {\bf 14}, L109 (1997)
gr-qc/9607071;

 R.~G.~Cai and Y.~Z.~Zhang,
  ``Black plane solutions in four-dimensional space-times,''
  Phys.\ Rev.\ D {\bf 54}, 4891 (1996)
  [gr-qc/9609065].
  
D.~Brill, J.~Louko and P.~Peld{\'a}n, 
Phys. Rev. {\bf D56}, 3600 (1997)
gr-qc/9705012;

L.~Vanzo,
``Black holes with unusual topology,"
Phys. Rev. {\bf D56}, 6475 (1997)
gr-qc/9705004;

D.~Birmingham,
  ``Topological black holes in anti-de Sitter space,''
  Class.\ Quant.\ Grav.\  {\bf 16}, 1197 (1999)
  [arXiv:hep-th/9808032].

D.~Birmingham and M.~Rinaldi,
  ``Brane world in a topological black hole bulk,''
  Mod.\ Phys.\ Lett.\  A {\bf 16}, 1887 (2001)
  [arXiv:hep-th/0106237].

  D.~Birmingham and S.~Mokhtari,
  ``Stability of Topological Black Holes,''
  Phys.\ Rev.\  D {\bf 76}, 124039 (2007)
  [arXiv:0709.2388 [hep-th]].
  
}


\baselineskip=15pt

\line{\hfill IFT UAM/CSIC-2015-103}

\vskip 0.7cm

\Title{\vbox{\baselineskip 12pt\hbox{}
 }}
{\vbox {\centerline{Holographic Complexity Of 
 }
\vskip10pt
\centerline{Cold Hyperbolic Black Holes}
}}

\vskip 0.5cm

\centerline{$\quad$ {\caps
Jos\'e L.F. Barb\'on
 and
Javier Mart\'{\i}n--Garc\'{\i}a
}}
\vskip0.7cm

\centerline{{\sl   Instituto de F\'{\i}sica Te\'orica IFT UAM/CSIC }}
\centerline{{\sl  Calle Nicolas Cabrera 13. 
 UAM, Cantoblanco 28049. Madrid, Spain }}
\centerline{{\tt jose.barbon@uam.es}}

\vskip0.7cm

\centerline{\bf ABSTRACT}

 \vskip 0.3cm

 \noindent
 AdS black holes with hyperbolic horizons provide strong-coupling descriptions of thermal CFT states  on hyperboloids. The low-temperature limit of these systems is peculiar.   In this note we show that, in addition to a large ground state degeneracy, these states also have an anomalously large holographic complexity, scaling logarithmically with the temperature. We speculate on whether this fact generalizes to other systems whose extreme infrared regime is formally controlled by  Conformal Quantum Mechanics, such as various instances of near-extremal charged black holes.  
 
 \vskip 1cm

\Date{October 2015}

\vfill

\vskip 0.1cm




\baselineskip=15pt

\newsec{Introduction}

\noindent

Computational complexity has been recently proposed as a new entry in the holographic dictionary between space-time geometry and quantum entanglements. Besides the total amount of entanglement, measured by the entanglement entropy and loosely associated to the `connectivity' of emergent space-time \refs{\eternalmalda,\mvr, \epr},  the sheer amount of emergent space seems to be associated to the degree of complexity of the entanglement structure, measured with respect to a set of elementary entanglement operations. 

One motivation for this correspondence arises form the tensor network models of geometry \refs{\tensors, \maldahartman}. In this picture, a notion of computational complexity of the state can be associated to the volume of the tensor network making it plausible that a quantitative volume/complexity relation may exist (cf.   \refs{\suscomple,\stanfordsus, \bridges, \enen,\llast}):
\eqn\ansatz{
C(t)\propto  {{\rm Vol}(\Sigma_t) \over G}
\;,}
where $\Sigma_t$ is a codimension-one space-like section  of the bulk with extremal volume, parametrized by some proper-time coordinate $t$. \foot{See \refs\lasttaka\ and \refs\action\ for further discussion of {\it ab initio} approaches to holographic complexity.}

A qualitative test of \ansatz\ is offered by the analysis of eternal AdS black holes, interpreted as thermofield-double states entangling two decoupled CFTs defined on spheres \refs{\eternalmalda, \maldahartman, \epr}. In this case, $t$ can be chosen as the standard AdS time in the global static frame, and the non-trivial time dependence of \ansatz\ comes from the portion of $\Sigma_t$ lying in the interior of the black hole. There is a saturating surface which represents a linearly growing wormhole section, so that one finds
 \eqn\rate{
 {dC \over dt} \sim T\,S\;,
 }
 for $t  \gg T^{-1}$, 
 where $T$ is the temperature and $S$ the entropy of the eternal black hole. If the wormhole is approximated by a flat tensor network, this formula can be interpreted as the growth of an effective quantum circuit of $S$ Qbits \refs{\maldahartman, \suscomple}, a representation which makes contact with the traditional definition of computational complexity in quantum information theory.\foot{The conditions for the tensor network to really represent a smooth wormhole geometry are not well understood \refs{\marpol, \icec, \noise}. }
    
The calculation of \rate\ within the eternal black hole geometry assumes implicitly that $T$ is sufficiently large to neglect finite-size effects. For standard AdS black holes, this means that $T\ell \gg 1$ where $\ell$  has the dual interpretation as the AdS$_{d+1}$ radius of curvature in the bulk and also the radius of the $(d-1)$-sphere where the $d$-dimensional CFTs are defined.   In this particular case the $T\,\ell \ll 1$ limit is on the other side of the Hawking--Page transition, and the $O(1/G)$ contribution to the complexity must be calculated in the vacuum AdS manifold, giving no contribution at this order to \rate. 

Alternatively, we can remove finite-size effects by working with black branes of non-compact horizon, where all integrated quantities, such as entropy and complexity, are extensive  in the CFT volume. In this case  we  implicitly refer to a  `complexity density'.\foot{This is in principle different from the strict definition of a notion of complexity for a subsystem, see \refs\alisha\ for discussions in this direction.} Black-brane metrics have the general form 
\eqn\bbrane{
ds^2 = -f(r) \,dt^2 + {dr^2 \over f(r)} + {r^2 \over \ell^2} \,dL^2
\;,}
where $dL^2$ stands for the spatial CFT metric and $f(r)$ has  a non-degenerate horizon at $r=r_0$ with Hawking temperature $T$, i.e. $f(r)\approx 4\pi T (r-r_0)$ near the horizon. We also require vacuum asymptotics  $f(r) \sim r^2 /\ell^2$ as $r\rightarrow \infty$.  

There are standard solutions given by 
\eqn\efes{
f(r) = k + {r^2 \over \ell^2} -{\mu \over r^{d-2}}
\;.}
 for flat $(k=0)$ and hyperbolic $(k=-1)$ CFT metrics. In the first case the usual UV/IR relation $r_0 \sim \ell^2 T$ holds down to zero temperature, with the entropy vanishing as $T^{d-1}$. In the second case, the CFT lives on a hyperboloid of curvature radius $\ell$. This system has exotic properties at low temperatures \refs\roberto, in particular a gross violation of the third law of thermodynamics 
 \eqn\zen{
 \lim_{T\to 0} S \longrightarrow S_0 = N_* \,V\,\ell^{1-d}\;,
 }
 where  $N_* \sim \ell^{d-1} /G \gg 1$ is the effective number of `species' in the strongly-coupled CFT. 
  
 The purpose of this note is to study some properties of  the holographic complexity, as defined by the {\it ansatz} \ansatz,  in such degenerate systems. In particular, we shall consider the concrete case of thermofield double states for pairs of CFTs on hyperboloids, as defined by AdS hyperbolic black holes \refs\hypobh. We begin in section 2 with a review of the relevant geometries and we continue in section 3 with the approximate calculation of the complexity.  
  
 \newsec{Cold Hyperbolic Horizons} 
 
 \noindent
 
We consider the black-brane metric \bbrane\ with hyperbolic horizon geometry, which has the interpretation of a thermal state for a CFT on a spatial $(d-1)$-dimensional hyperboloid H$_{d-1}$:
\eqn\hu{
f(r) = -1 +r^2  - {m \over r^{d-2}} \;, \qquad dL^2 = d{\rm H}_{d-1}^2 = d\chi^2 +  \sinh^2 (\chi) \,d\Omega_{d-2}\;,
}
where we measure lengths in units of the curvature radius $\ell =1$. Alternatively, the maximally extended geometry can be interpreted, following \refs\eternalmalda,  as dual to a thermofield double state on the direct product of two copies of the CFT on respective hyperboloids. 

The mass parameter $m$ is related to the horizon radius $r_0$ by
\eqn\mup{
m = r_0^{d-2} (r_0^2 -1)
\;,}
and the Hawking temperature is given by
\eqn\h{
T = {r_0 \over 4\pi} \left(d - {d-2 \over r_0^2}\right)
\;.}
The case of vanishing mass parameter is special, corresponding to $T = 1/2\pi$, the Rindler temperature. At this particular value the metric is nothing but a Rindler patch of the  vacuum AdS manifold. The corresponding thermofield double on two decoupled hyperboloids is conformally equivalent to the hemispherical decomposition of a single copy of the CFT on a unit sphere (cf. for example \refs\mvr). The same bulk geometries can also be interpreted as computing properties of a certain entangled state on two static patches of a de Sitter CFT (cf. for example \refs\mukund).

Here, we are more interested in the  $T\rightarrow 0$ limit, where the horizon drops to a minimum radius
\eqn\minrad{
r_c = \sqrt{d-2 \over d}
\;,}
 corresponding to a negative mass parameter
\eqn\muc{
m_c = -{2 \over d} \left({d-2 \over d}\right)^{d-2 \over 2}
\;.}
In this extremal case the function $f(r)$ develops a double zero at the horizon, namely in the vicinity of $r=r_c$ we can write
$$
f(r)_{T=0} = d\cdot (r-r_c)^2 + \dots
$$
where the dots stand for terms of order $(r-r_c)^3$ or higher. This suggests that we can parametrize the low-temperature geometries in terms of the radial variable $\rho= r-r_c$. Then, to first non-trivial order in $\rho$ and $\rho_0 = r_0 - r_c$ we have 
$$
f(r) \approx d\cdot (\rho^2 - \rho_0^2) + \dots\;, 
$$
an approximation good for $\rho_0 \leq \rho \ll r_c$. The low-temperature horizon sits at $\rho=\rho_0 \approx 2\pi T /d  +O(T^2)$.  

The black AdS geometry changes character in the vicinity of the minimum radius $r\sim r_c$, so that the region 
$\rho_0 \ll \rho \ll r_c$, arising at very low temperatures,  is approximately described by ${\rm AdS}_{1+1} \times {\rm H}_{d-1}$, i.e. the hyperbolic `space' decouples from an asymptotic AdS$_2$ factor.  The corresponding curvature radii are given by 
$$
\ell_{{\rm AdS}_2} = {1\over \sqrt{d}}\;, \qquad \ell_{{\rm H}_{\rm IR}} = \sqrt{d-2 \over d}\;,
$$
measured in units $\ell =1$. We will refer to this factorized geometry as the CQM region, to signify the formal AdS$_2$/CFT$_1$ duality to some hypothetical  Conformal Quantum Mechanical (CQM) system that would describe the deep infrared regime.
 
\newsec{Estimating Holographic Complexity}

\noindent

Following the formula   \ansatz, we compute the holographic complexity as the volume of extremal codimension-one surfaces in the given geometry, parametrized by the static asymptotic time variable. We shall also parametrize the absolute normalization of \ansatz\ as differing from the RT formula of entanglement entropy \refs\rt\ by a factor $\alpha$.   

The exact variational problem is complicated, but a useful order-of-magnitude estimate can be obtained by an approximate description of the full metric \bbrane, according to a piece-wise approximation for the function $f(r)$. For $r\gg r_0$ we can approximate the metric by the vacuum AdS$_{d+1}$ solution. In the near-horizon region $r_0 < r< r_R$, with $r_R$ an $O(1)$ multiple of $r_0$, we can take the Rindler approximation, whereby the metric is expressed as a product of two-dimensional flat space and the horizon:
\eqn\rme{
ds^2_{\rm Rindler} \approx -(dX^0)^2 + (dX^1)^2 + r_0^2  \;d{\rm H}_{d-1}^2
\;,}
where
\eqn\cha{
X^0 = \sqrt{r-r_0 \over \pi T} \,\sinh(2\pi T t)\;, \qquad X^1 = \sqrt{r-r_0 \over \pi T} \,\cosh(2\pi T t)\;,
}
a change of variables valid for $r>r_0$ on one of the asymptotic regions. 
Finally, the interior geometry is parametrized in Schwarzschild coordinates $(r,t)$, formally continued to $r<r_0$, with $r$ now denoting a time-like coordinate and $t$ a space-like one. There is an analogous extension of the Rindler patch to the interior, with the analogous  change of variables 
\eqn\chain{
 X^0 = \sqrt{r_0-r \over \pi T} \,\cosh(2\pi T t)\;, \qquad X^1 = \sqrt{r_0-r \over \pi T} \,\sinh(2\pi T t)\;.
}

Within this prescription we view the portion of the extremal surface lying outside the horizon as composed of two pieces: an asymptotic component $\Sigma_{\rm UV}$ which is well approximated by a constant $t$ surface in AdS$_{d+2}$, and  a `Rindler piece' $\Sigma_{\rm R}$, parametrized by
a curve on the $(X^0, X^1) $ plane of  \rme. Within the Rindler patch, local volume for fixed $X^1$ interval is maximized by the $X^0 = {\rm constant}$ surfaces, and thus we take this ansatz for $\Sigma_{\rm R}$. 
For $t=0$, this is all there is, since the extremal surface is just the $t=0$ section of the extended geometry, with the two exterior geometries   glued by the horizon. However, as $t$ grows, the surface enters the horizon at higher values of $X^0$ and tends to extend through the interior patch of the black brane geometry. The $X^0 = {\rm constant}$ ansatz continues to be reasonable as long as the  complete surface stays inside the  interior Rindler region. Since the only length scale controlling the width of the Rindler region is $T^{-1}$, the approximation $X^0 = {\rm constant}$  must break down for large times, $t \gg T^{-1}$. 

\bigskip
\centerline{\epsfxsize=0.3\hsize\epsfbox{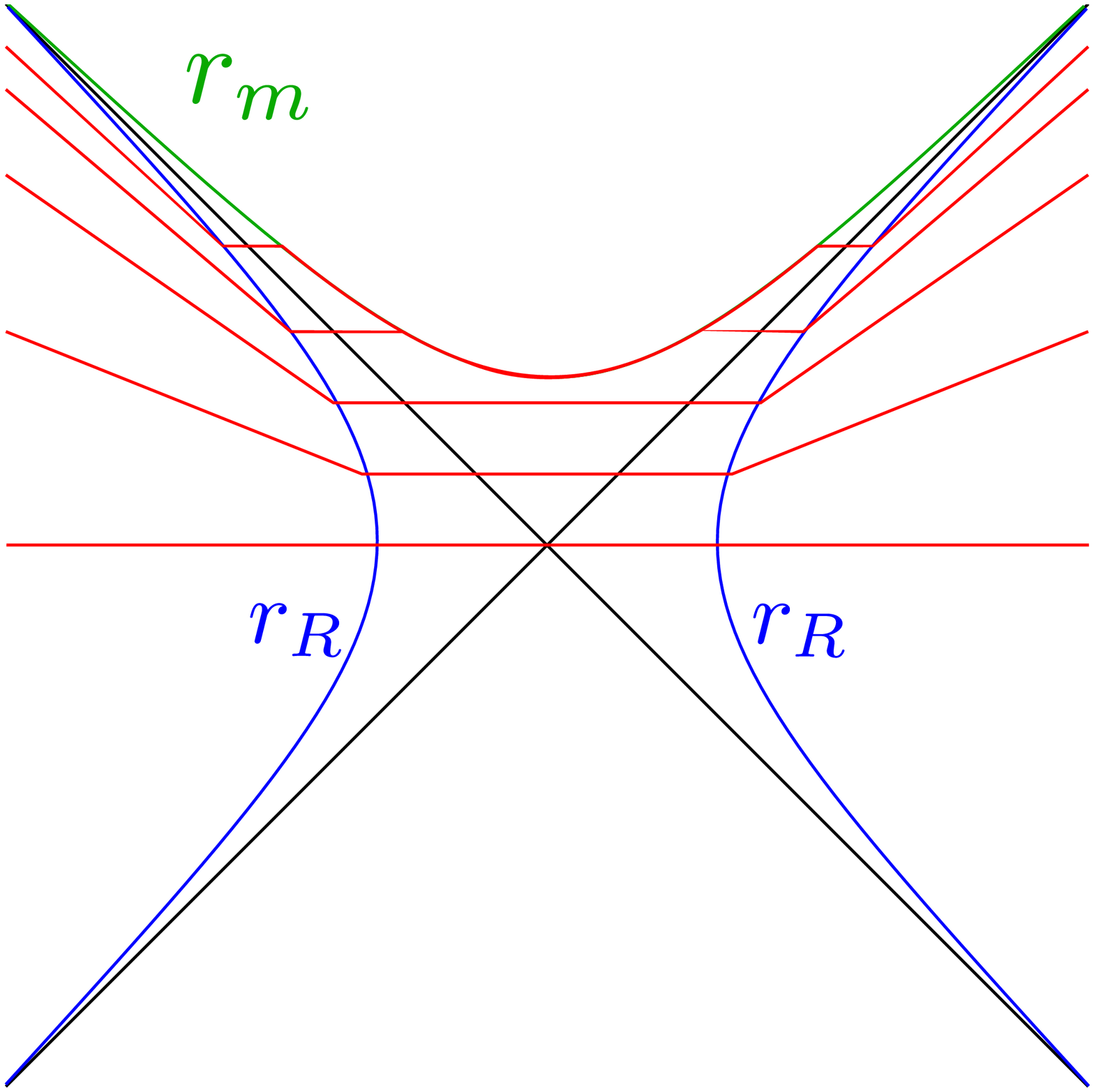}}
\noindent{\ninepoint\sl \baselineskip=2pt {\bf Figure 1:} {\ninerm
Piecewise decomposition of $\Sigma_t$, represented by space like sections at different times. For generic times, straight tilted segments correspond to $\Sigma_{\rm UV}$, horizontal segments give  $\Sigma_{\rm R} $ and hyperbolic segments on the $r=r_m$ surface correspond to $\Sigma_{\rm WH}$. The $t=0$ surface lacks an interior component. As $t$ increases from zero, an interior component  begins to develop gradually, as $\Sigma_{\rm R}$ eventually transmutes into $\Sigma_{R} \cup \Sigma_{\rm WH}$.   }}
\bigskip
At very long times, there is a natural answer for the variational problem in the interior, since the surfaces $r={\rm constant}$ are invariant under the $t$-translation isometry. The volume of a $\Delta t$ portion of such $r={\rm constant}$ surfaces is proportional to  
$$
\Delta t\, r^{d-1} \sqrt{|f(r)|}\;,
$$
so that stationary points $r_m$  of this function determine extremal surfaces far from the `exit point', i.e. for  large $\Delta t$. In all cases studied in this paper, one finds $|r_0 - r_m | \sim |r_R - r_0 |$, implying that $r_m$ is always close to the inner edge of the Rindler region and, in particular, it is roughly symmetrical of the $r=r_R$ surface by a reflection through the horizon (see figure 1). As a consequence, the `exit point' from the $r = r_m$ surface is approximately given by $t_{\rm exit} \approx t$, where $t$ is the time label of the exterior asymptotic surface $\Sigma_{\rm UV}$.

 The approximate ansatz for the extremal surface is thus  $\Sigma_{\rm WH} \cup \Sigma_{\rm R} \cup \Sigma_{\rm UV}$, where $\Sigma_{\rm WH}$ is the $r=r_m$ surface along the `wormhole' in the interior, cut off at $t_{\rm exit} \sim t$, with total $t$-length  of order $\Delta t \sim 2t$.

Within this construction, the volume of $\Sigma_{\rm UV}$ is independent of $t$, whereas the volume of $\Sigma_{\rm R}$ vanishes at large $t$, being delimited by two curves (interior and exterior) asymptotic to the same horizon. Therefore, the rate of growth of the complexity is controlled by $\Sigma_{\rm WH}$ at large times. A graphical representation of the piecewise decomposition of $\Sigma_t$ is shown in figure 1.

\subsec{High Temperatures}

\noindent 

At high temperatures  $T\gg 1$ we have  $f(r) \approx r^2 - r_0^d /r^{d-2}$ and $r_0 \approx 4\pi T /d$. 
Evaluating the volume of $\Sigma_t$, we find the standard result \rate\ for the long-time growth rate, with $S\sim N_* V T^{d-1}$ the high temperature entropy of the large-$N_*$ CFT on the hyperboloid.  At $t=0$, we can distinguish two qualitatively different contributions. First we have the  UV contribution of $\Sigma_{\rm UV}$, 
  \eqn\uvc{
C_{\rm UV} = 2\alpha {V \over 4 G} \int_{r_R}^{r_\Lambda} {dr\,r^{d-1} \over \sqrt{f(r)}} \sim N_* \,V \, \left(\Lambda^{d-1} - T^{d-1} \right)\;,
}
where we can as well neglect the $T$-dependent term coming from the lower limit of the integral, since we are assuming $\Lambda \gg T$.
Second, we have a threshold contribution coming form $\Sigma_{\rm R}$:
\eqn\ritz{
C_{\rm R} \big |_{t=0} = 2\alpha {V \over 4 G} \int_{r_0}^{r_R} {dr\,r^{d-1} \over \sqrt{f(r)}} \sim N_* \, V\,T^{d-1} \sim S\;,
}
where we have used the Rindler approximation to the metric to estimate the integral in order of magnitude. In this expression, as well as others that follow, the matching ambiguity coming from the precise location of $r_R$ and the various errors from the piecewise matchings  of $\Sigma_t$ can be estimated by shifting $r_R$ an amount of $O(1)$, resulting in  an additive ambiguity of order $S$  for $C_{\rm R}$.  

The UV contribution to the complexity is constant in time. Denoting the rest of the complexity by $\Delta C (t) = C(t) - C_{\rm UV}$ we find the following behavior at high temperatures: 
\eqn\cit{
\Delta C(t) =O(S) \;\;{\rm for}\;\;t < T^{-1}\;, \qquad \Delta C(t) \sim S\,T\,t\;\;{\rm for} \;\;t\gg T^{-1}
\;.}

\bigskip
\centerline{\epsfxsize=0.4\hsize\epsfbox{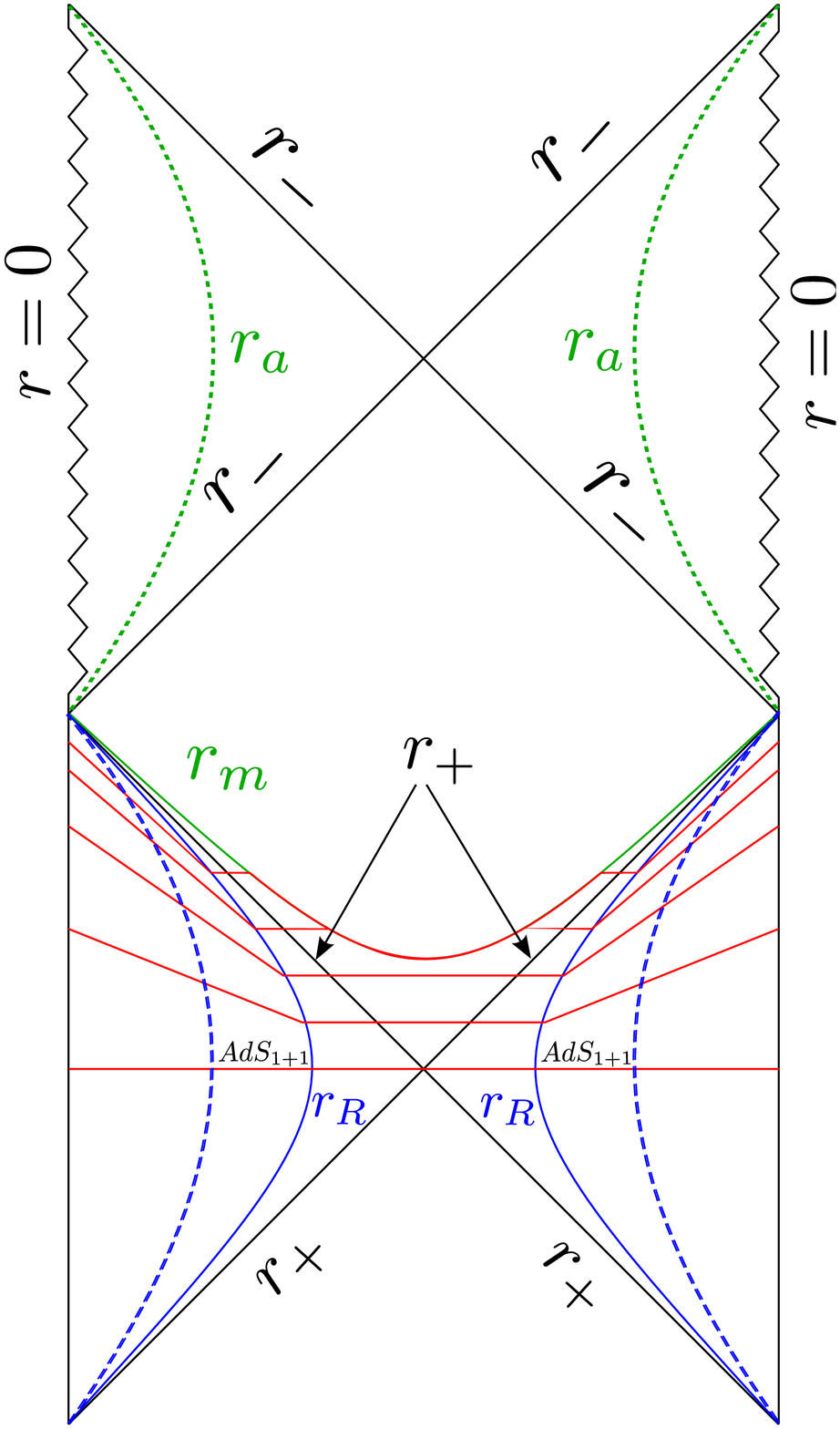}}
\noindent{\ninepoint\sl \baselineskip=2pt {\bf Figure 2:} {\ninerm
The global structure of the hyperbolic black hole interior is similar to that of charged black holes. There is an inner horizon $r_-$ which separates the space-like and time-like character of fixed-$r$ surfaces in the interior. For instance, one has the inequalities $0<r_a < r_- < r_m < r_+ = r_0 < r_R$. The intermediate AdS$_2$ region develops only for low temperatures. }}
\bigskip

\subsec{Low Temperatures}
\noindent

 Our main interest is the low-temperature regime, $T\ll 1$ in $\ell=1$ units, where the thermodynamics becomes more exotic. In evaluating the volume of extremal surfaces, we must distinguish the qualitatively different regions of the bulk geometry, namely for $r\gg r_c$ we have an approximately AdS$_{d+1}$ geometry with a time slicing adapted to the ${\bf R} \times {\rm H}_{d-1}$ CFT frame, and for $\rho_0 \ll \rho\ll r_c$ 
we have a ${\rm AdS}_{1+1} \times {\rm H}_{d-1}$ geometry. Accordingly, the codimension-one  surfaces split as  (see figure 2) 
$$
\Sigma_t \sim \Sigma_{\rm WH} \cup \Sigma_{\rm R} \cup \Sigma_{\rm CQM} \cup \Sigma_{\rm UV}
\;.
$$
Here $\Sigma_{\rm UV}$ extends for $r\gg r_c$. The new portion extending   along the AdS$_{1+1}$ radial slice $\rho_0 \ll \rho\ll r_c$
will be denoted  $\Sigma_{\rm CQM}$.  Finally, in the deep infrared region we have the Rindler portion $\Sigma_{\rm R}$ given by the interval $\rho_0 < \rho < \rho_R$, with $\rho_R$ an $O(1)$ multiple of $\rho_0$.  In the interior we find  the wormhole portion $\Sigma_{\rm WH}$ along $r=r_m$. For all partitions except $\Sigma_{\rm UV}$ we can regard the hyperbolic H$_{d-1}$ factor as an spectator. 

We first discuss the situation at $t=0$, where $\Sigma_{\rm WH}$ is absent. The contribution from $\Sigma_{\rm UV}$ is the standard 
$N_* V \Lambda^{d-1}$. The contribution from $\Sigma_{\rm CQM}$ is interesting because the complexity picks equal contributions for every region of the CQM region,
\eqn\cqm{
C_{\rm CQM} (0) \approx 2 \alpha {r_c^{d-1} V \over 4G} \int_{\rho_R}^{r_c} {1\over \sqrt{d}}{d\rho \over  \rho}  = {2\alpha \over \sqrt{d}} \,S_0 \, \log(1/T)\;,
}
leading to a logarithm with a characteristic coefficient controlled by the zero-temperature entropy of the system. The $O(1)$ ambiguities at the endpoints of the integral amount to an additive error of order $S_0$. Notice however that the coefficient of the logarithm, given by $2\alpha S_0 /\sqrt{d}$,  is robust in the low $T$ limit.

Finally, the Rindler contribution coming from $\Sigma_{\rm R}$ is of order
\eqn\rtzero{
C_{\rm R} (0) \approx 2 \alpha {r_c^{d-1} V \over 4G} \int_{\rho_0}^{\rho_R} {1 \over \sqrt{d}} {d\rho \over \sqrt{\rho^2 - \rho_0^2} }\sim S_0\;,
}
where the matching errors are also of order $S_0$.

As before, the exterior surfaces  in both the CQM and UV regions have a time-independent volume. Hence the time development of the complexity proceeds by the gradual deformation of $\Sigma_{\rm R}$ into $\Sigma_{\rm WH} \cup \Sigma_{\rm R}$. As can be seen from figure 2, the volume of $\Sigma_{\rm R}$ is negligible at large times, whereas that of $\Sigma_{\rm WH}$ is controlled by the local maximum of $r^{d-1} \sqrt{|f(r)|}$. Since we are working at very low temperatures,  it is tempting to pick the $O(1)$   radius $r=r_a$ which maximizes  the $T=0$ function
$$
r^{d-1} \sqrt{\left|1-r^2 + {m_c \over r^{d-2}}\right|}\;
.$$
However, there is a subtlety. This $O(1)$ maximum at $r=r_a$ survives for small but non-zero $T$, but in fact we have $f(r_a) >0$, implying that $r=r_a$ is a time-like surface (shown in figure 2). It turns out that there is a small $T$-dependent local maximum of $r^{d-1} \sqrt{|f(r)|}$, with height of $O(T)$,  within the interior Rindler region (see figure 3). The corresponding $r=r_m$ surface is space-like, since  $f(r_m) <0$. In this regime the function to be maximized  is approximately given by
$$
(r_c + \rho)^{d-1} \sqrt{d|\rho_0^2 - \rho^2|}\;,
$$
which is maximized close to $\rho_m =0$, so that the WH surface is given by $\rho \approx 0$. Again, it is roughly the symmetrical  of  the $\rho=\rho_R$ surface by a reflection with respect to the horizon, implying that $t_{\rm exit} \sim  t$ and thus  a 'wormhole length' of order $\Delta t \approx 2t$. The resulting   large $t$ complexity is
\eqn\whiii{
C_{\rm WH} (t) =2\,t\, \alpha\, {V r_c^{d-1} \over4 G} \sqrt{d} \,\rho_0 \approx \alpha {4\pi T \over \sqrt{d}} \,S_0 \, t\;.}
Grouping together these results and restoring the curvature radius $\ell$, we find a total low-temperature subtracted complexity  given by
\eqn\fins{
\Delta C(t) \approx {2\alpha \over \sqrt{d}} \,S_0 \, \log (1/\ell \,T) \;,\;{\rm for} \;\;t< T^{-1}\;,}
at small times and 
\eqn\lod{
 \Delta C(t) \approx {2\alpha \over \sqrt{d}} \,S_0 \, \log (1/\ell \,T)
+ {4\pi \alpha \over \sqrt{d}} \,S_0 \, T\,t \;\;{\rm for}\;\;t\gg T^{-1}\;.
}
at long times. It should be noted that, while we have kept the coefficient found in \whiii, it must be understood as an estimate with 
$O(S_0)$ additive ambiguities, unlike the coefficient of the time-independent logarithmic term, which is a robust prediction for the strongly coupled CFT. 

\bigskip
\centerline{\epsfxsize=0.5\hsize\epsfbox{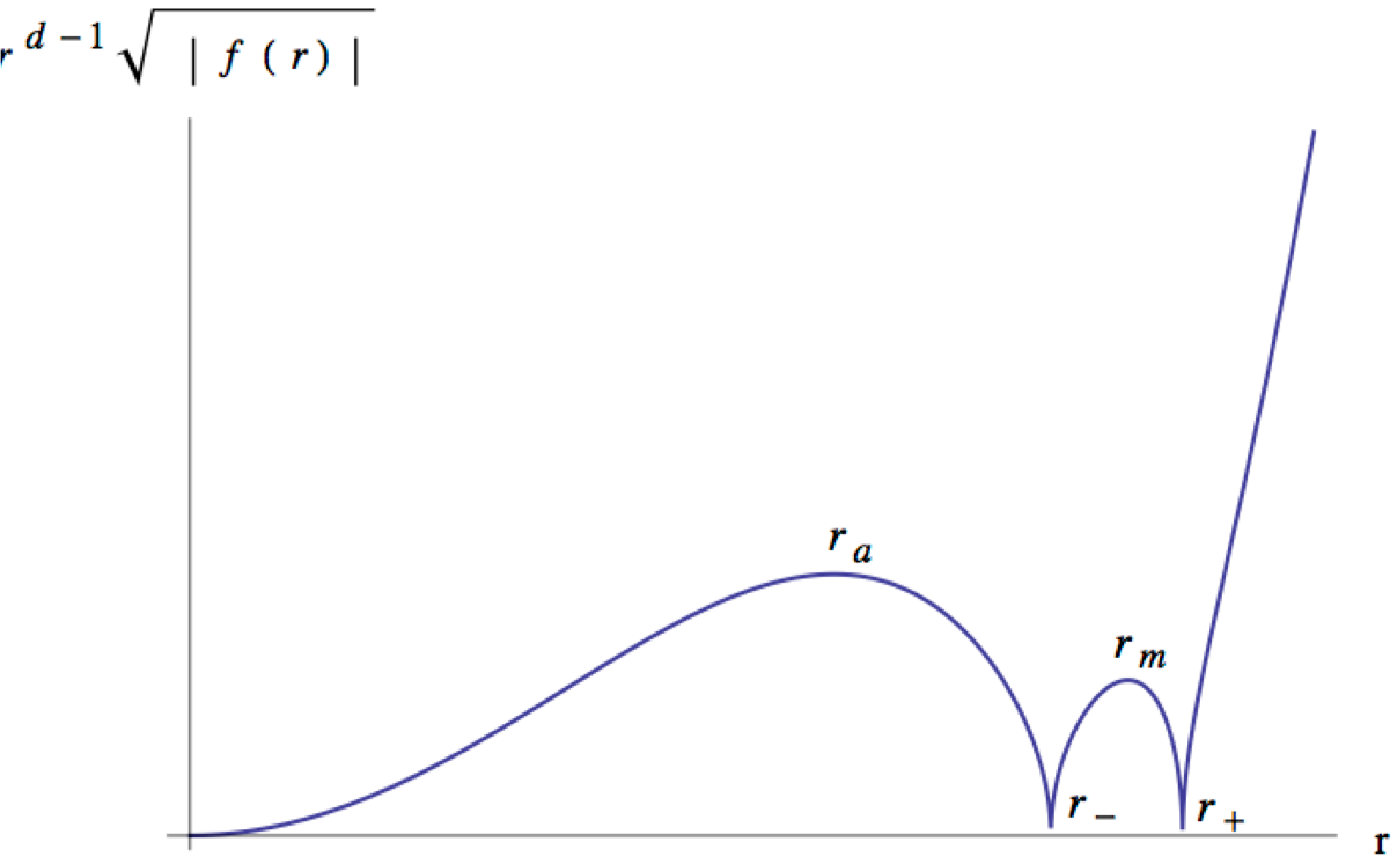}}
\noindent{\ninepoint\sl \baselineskip=2pt {\bf Figure 3:} {\ninerm
The function $r^{d-1} \sqrt{|f(r)|}$ at low temperatures, showing the small maximum of $O(T)$ at $r=r_m$ in the near-horizon (Rindler) region, and the
$O(1)$ maximum at $r=r_a$. In the high-$T$ regime the Rindler bump grows larger than the local maximum at $r=r_a$, which becomes a small detail near the singularity.   }}
\bigskip

\newsec{Conclusions}

\noindent

We have studied the structure of low-temperature thermofield double states in strongly coupled CFTs defined on hyperboloids. In particular, we have focused on properties characterized by the gravitational description in terms of AdS hyperbolic black holes. 
We have found that, in addition to the known zero-temperature entropy of order $N_*$, these states have a large holographic complexity, as measured by extremal bulk volumes. 

The low-temperature geometry develops a AdS$_2$ throat contributing a time-independent complexity of order
\eqn\troat{
C_{\rm CQM} = {2\alpha S_0 \over \sqrt{d}} \,\log (1/T\,\ell) + O(S_0) \;,
}
where $\ell$ is the radius of curvature of the hyperboloid and $S_0$ is the zero-temperature limit of the entropy. This expression was derived up to additive ambiguities of  $O(S_0)$, but the coefficient of the logarithm is a  reliable strong-coupling prediction in the low-$T$ limit, once we define the absolute normalization of the complexity, which entails fixing the value of $\alpha$.

This means that tensor network representations of these states  should incorporate this logarithmic tube. It would be interesting to find an explicit tensor model realizing this fact. Perhaps these ideas can be checked at weak coupling, searching for vestiges of the large low-temperature complexity directly in the perturbative field theory wave functions.   

Since the behavior \troat\  is controlled by the emergence of the AdS$_2$ throat, it is tempting to take it at face value, as a general property of any near-extremal geometry of Reissner--Nordstrom type. This includes the benchmark model of AdS/CMT, the near extremal charged AdS$_4$ black brane, with either chemical potential $\mu$ or magnetic field $B$ (cf. \refs\rcmt\ for a review). In those systems, the same expression \troat\ follows,
with the substitution  of the curvature scale $1/\ell$ by an effective mass of the order of $\mu$ and/or $\sqrt{B}$ (cf. \refs\fuertes\ for a discussion of peculiar properties of entanglement entropy in these systems). On the other hand, it is also known that consistent
string theory embeddings of these finite-density systems tend to show perturbative instabilities, the near extremal black holes being unstable to the condensation of clouds of classical charged hair (see \refs\action\ for a discussion of possible relations to complexity). 

The low-temperature hyperbolic black holes studied in this note show no sign of any perturbative instability when embedded in string theory, but they are not free from potential  non-perturbative instabilities. As shown in \refs\magan,  if these  black holes carry $N$ units of RR charge, they can super-radiate it spontaneously, causing the fragmentation of the black hole by brane emission (see also \refs{\insightfull,\craps, \usu, \har}). This process was studied  in \refs\magan\ for compact hyperbolic horizons, but it may take place in the present non-compact set up by nucleation of critical bubbles just hovering above the horizon, leading to a sort of condensation of RR hair. It would be interesting to further clarify these issues and their possible impact on the prediction \troat.

\bigskip{\bf Acknowledgements:} 

We thank E. Rabinovici for discussions. This work   was partially supported by MINECO and FEDER under a grant  FPA2012-32828, and the 
spanish MINECO {\it Centro de Excelencia Severo Ochoa Program} under grant SEV-2012-0249.

{\ninerm{
\listrefs
}}

\end

\newsec{Hyperbolic Black Holes As Degenerate Systems}

\noindent

There are a number of strategies to engineer bulk cosmological singularities in AdS/CFT models of crunch/bang singularities. One simple implementation is to start with a CFT on a singular frame, i.e. the non-dynamical metric where the CFT is defined is itself a singular cosmology, and study how this singularity is realized by the bulk dynamics. Among the infinite set of models of this type, one looks for those whose dual bulk dynamics is relatively easy to construct.

A simple example of this sort is provided by  the Kasner metrics 
\eqn\ks{
ds^2_{\rm CFT} = -dt^2 + \sum_{i=1}^{d-1}  t^{2p_i} \,d x_i^{\,2} \;,}
for generic parameters $p_i$ satisfying $\sum p_i =\sum_i p_i^2 = 1$. Despite its anisotropic character, this frame has the  technical advantage of being Ricci flat, which allows us to write one bulk solution with no extra work, namely  
\eqn\kb{
ds^2 = {dr^2 \over r^2} + r^2 \left( -dt^2 + \sum_{i=1}^{d-1}  t^{2p_i} \,d x_i^{\,2}\right)\;.} 

This solution provides a large-$N$ definition of a certain CFT state on the Kasner frame, which we refer to as the `Kasner state'.
Its global structure resembles an AdS Poincar\'e patch, cut by the singularity $t=0$ on the time-reflection spatial surface.

Another model with a simple bulk dual is a frame of the form ${\rm E}_d \times {\bf S}_{R(t)}^1$, where ${\rm E}_d = {\bf R} \times {\bf S}^{d-1}$ is the standard Einstein static universe and the extra ${\bf S}^1$ has a time-dependent radius,  $R(t)$, which is  carefully chosen so that the bulk solution admits a simple parametrization. In particular, working in units of the $(d-1)$-sphere radius, the choice 
\eqn\scc{
ds^2_{\rm CFT} = -dt^2 + d\Omega_{d-1}^2 +  \cos^2 (t) \,d\phi^2\;,
}
with the periodicity  $\phi \equiv \phi + 2\pi R $, arises as a conformal  boundary of a periodically identified AdS$_{d+2}$ manifold.  To see this, it is convenient to map the CFT frame by a conformal transformation to ${\rm dS}_d \times {\bf S}^1_R$, where we transfer the time dependence from the circle to a rescaling of the E$_d$ factor, leading to a de Sitter cosmological frame times a fixed circle,
$$
ds^2_{\rm CFT} = \cos^2 (t)\left[-d\tau^2 + \cosh^2(\tau) \,d\Omega_{d-1}^2 + d\phi^2 \right]\;.
$$
Here we require $dt= d\tau /\cosh(\tau)$, which fixes the time-diffeomorphism $\cosh(\tau) = 1/\cos(t)$. The metric in square brackets, ${\rm dS}_d \times {\bf S}^1$,
is a conformal boundary metric of
\eqn\bscc{
ds^2_{d+2} = \cosh^2 (\rho) \,d\phi^2 + d\rho^2 + \sinh^2 (\rho) \left(-d\tau^2 + \cosh^2 (\tau)\,d\Omega_{d-1}^2 \right)
\;,}
which is locally the AdS$_{d+2}$ vacuum, up to the identification of the $\phi$ coordinate. \foot{To see this, notice that Wick rotation in the $\tau$ time variable gives a standard representation of Euclidean AdS$_{d+2}$.} The compact nature of $\phi$ produces a global crunch
in the bulk, as seen by continuing \bscc\ past the horizon at $\rho=0$. We can do this by the standard continuation $\rho =i\eta$ and $\tau = \chi -i\pi/2$ which produces a FRW form of the metric:
\eqn\frwc{
ds^2 = -d\eta^2 + \sin^2 (\eta)\left(d\chi^2 + \sinh^2(\chi) d\Omega_{d-1}^2 \right) + \cos^2 (\eta)\,d\phi^2\;,
}
consisting on a circle of radius $R\cos(\eta)$ fibered over the FRW patch of vacuum AdS$_{d+1}$. A singularity arises at $\eta=\pi/2$ where
the circle shrinks to zero size. Since the bulk metric is locally AdS with identifications, we shall refer to this model as the `topological crunch'. This model was studied in \refs\bana\ and recently analyzed from the point of view of holographic entanglement entropy in \refs\maldads.

If we look at this model in the de Sitter frame, a CFT on ${\rm dS}_d \times {\bf S}^1$ can be viewed as inducing a non-conformal theory on dS$_d$ with mass scale $M=1/2\pi R$. This suggests that there should exist versions of this set up involving massive deformations of CFTs on $d$-dimensional de Sitter space. If we introduce the mass scale $M$ via a relevant operator, the UV of the field theory is still $d$ dimensional and we are led to a class of extensively studied models of AdS crunches (cf.  for example \refs{\horo, \usu, \har}). 

The starting point of these constructions is a CFT on dS$_d$, perturbed by a relevant operator $\CO$ of weight $\Delta <d$, 
\eqn\relv{
\delta I = \int_{{\rm dS}_d} M^{d-\Delta} \;\CO
}
 where  $M$ is a fixed mass scale, expressed in units of the Hubble parameter. In terms of standard QFT intuition, the effect of this operator is clearest in the limit $M\gg 1$, since in this case the massive deformation decouples from the Hubble expansion. Although generically we may expect a gapped theory at the scale $M$, we may also  have a non-trivial infrared CFT surviving at distances much larger than $M^{-1}$, which is subsequently placed on an slowly expanding de Sitter space-time. The bulk picture for this scenario is a de Sitter-invariant  background 
 \eqn\sds{
 ds^2_{\rm bulk}  = d\rho^2 + f(\rho)^2 \,ds^2_{{\rm dS}_d}\,
}
  with the warp function $f(\rho)$ depending on $M$. When $M\gg 1$ we expect this function to be well approximated by a narrow domain wall, separating two almost-vacuum AdS backgrounds $f_\pm (\rho) = \ell_\pm \sinh(\rho/\ell_\pm)$ with slightly different curvatures. We take the asymptotic (UV) AdS, denoted 
AdS$_+$, to have radius of curvature $\ell_+ =1$, whereas the infrared one, AdS$_-$, has radius $\ell_- < 1$. Accordingly, standard renormalization-group intuition implies $N_+ > N_-$ for the effective species numbers, $N_\pm = \ell_\pm^{\,d-1} /G$.  

If we now pick a global coordinate system for the asymptotic AdS$_+$ patch, with metric 
$$
ds^2_{{\rm AdS}_+} \approx -(1+r^2)\,dt^2 + {dr^2 \over 1+r^2} + r^2\,d\Omega_{d-1}^2\;,
$$
the domain wall at $\rho=\rho_M$ will execute a de Sitter-invariant motion of the form 
\eqn\te{
r(t)_{\rm wall} = \left({1+r_M^2 \over \cos^2 (t)} -1\right)^{1/2}\;,
}
in the global coordinate system, where $r_M = \sinh(\rho_M)$ is  interpreted as  the minimum value of $r(t)$. Since the global coordinate system is adapted to the static frame of the CFT, the background \sds\  can be interpreted as a $t$-dependent state on the same CFT defined on the Einstein universe (E-frame) 
\eqn\eu{
ds^2_{{\rm E}_d} = -dt^2 + d\Omega_{d-1}^2\;.}
For $M\gg 1$ we have the UV/IR relation $\sinh (\rho_M) =r_M \approx M$. 
Moreover, at any time such that $r(t)\gg 1$ we can approximate \te\ by  $r(t) \approx r_M /\cos (t)$. Defining a $t$-dependent version of the UV/IR relation in the Einstein frame,  $M(t) = r(t)$, we learn that this background can be interpreted as a deformation of the E-frame CFT by the same relevant operator, but now with a $t$-dependent mass scale 
\eqn\ms{
M(t) = {M \over \cos(t)}\;.}
Of course, we could have anticipated this result by noticing that the dS and E frames of the CFT are conformally related, 
$
ds^2_{{\rm dS}_d} = \cosh^2 (\tau) \,ds^2_{{\rm E}_d}$, 
provided we link the time variables according to 
$
dt  = d\tau / \cosh (\tau)
$. So the $t$-dependent mass scale \ms\ arises as the simple consequence of rewriting the relevant perturbation \relv\ in the Einstein frame:
\eqn\relve{
\delta I = \int_{{\rm E}_d} (M(t))^{d-\Delta} \,\CO\;.
}

In the case that the relevant operator leaves no large-$N$ CFT in the infrared, we can expect an  interior  `non-geometry' of stringy curvature or, more neatly, a `bubble of nothing'  \refs\wbn, which effectively cuts off the bulk at $\rho=\rho_M$ (see \refs\mukundmar\ and references therein). Equivalently, the bubble of nothing grows in global coordinates following \te\ above. 

In the case when the perturbation is softer than Hubble,  $M\ll 1$, the effects of the expansion cannot be easily disentangled from those of the relevant operator and the QFT intuition is less clear. In the holographic picture we can still consider the natural extrapolations of these dS-bubble backgrounds:  either bubbles of nothing with minimal radius much smaller than Hubble, or small bubbles with non-trivial interior.  In the second case, since the bubble wall has an acceleration horizon, we can continue the geometry past it,  using a standard parametrization which preserves the de Sitter isometry group, i.e. a hyperbolic FRW model: 
\eqn\buint{
ds^2_{\rm interior} = -d\eta^2 + a(\eta)^2 \left(d\chi^2 + \sinh^2 (\chi)\,d\Omega_{d-1}^2 \right)\;,
} 
with $a(\eta)$ a scale factor satisfying $a(\eta) \approx \eta$ near $\eta=0$ in order to match smoothly at the acceleration horizon of the bubble. In general, the complete solution will require specifying the matter fields $\Phi$ which are dual to the relevant CFT operator $\CO$, and  contribute an extra energy density on top of the AdS vacuum. At any rate,  generic FRW models with negative cosmological constant end with a crunch occurring at a zero of the scale factor $a(\eta_\star)=0$. 
Hence, the hyperbolic slices expand initially, reach a maximum volume with scale factor  $a_m = a(\eta_m)$ at some intermediate time $0<\eta_m<\eta_\star$, and collapse back to zero volume at the crunch time $\eta_\star$. 
The $M\gg 1$ model with a  thin domain wall  corresponds to the solution $a(\eta) \approx \sin(\eta)$, which would remain a good approximation up to the vicinity of the crunch at $\eta_\star \approx \pi$. 

In either case, independently of the size of $M$ compared to Hubble,  the crunch singularity acquires the QFT interpretation of a  singularity of the driving by the E-frame  relevant operator, characterized by a $t$-dependent mass scale $M(t)$ having a pole at $t=t_\star$. Conversely, the dS-frame description of the same state reparametrizes the finite $t$ interval up to $t_\star$ into the `eternity' of dS space-time giving a completely non-singular description of the physics.  It is in this sense that we can regard the dS-frame description as a `holographic definition' of the crunch singularity (cf. \refs{\maldab, \harlowsus, \falls, \complmaps}). We have collected the different qualitative features of the dS deformed models in Figure 1.

\bigskip
\centerline{\epsfxsize=0.4\hsize\epsfbox{penroseNearextremal2.eps}}
\noindent{\ninepoint\sl \baselineskip=2pt {\bf Figure 1:} {\ninerm
Global structure of the bulk space-time  dual to massive deformations of dS CFTs. All three space-times have vacuum AdS on the exterior of a bubble whose wall sits at constant $\rho_M$. On the left we have a dS-invariant bubble with nothing inside, dual to a gapped phase. In the center we have a dS-invariant bubble with an approximate AdS-FRW interior and  thin walls.  On the right we have a smaller dS-invariant bubble with `thick' walls and a more generic negatively curved FRW interior. All these solutions have $O(1,d)$ isometries, and can be obtained as the Lorentzian continuation of Euclidean solutions with AdS asymptotics and $O(d+1)$ symmetry.  }}
\bigskip

\newsec{Complexity Estimates}

\noindent

In evaluating 
the {\it ansatz} \ansatz\  for the models at hand we parametrize the extremal surfaces as a function of boundary time coordinates  $t$ for which the singularity is `finitely' far away. In this way we attempt to capture the notion of `complexity of the singularity'. We shall take the models of the previous section in backward order, starting with the dS crunch models.

\subsec{Complexity of dS-CFT Crunches}
\noindent

 Since all these backgrounds are asymptotically AdS near the boundary, there is a piece of the extremal surface $\Sigma_{\rm UV}$, which is well approximated by a $t={\rm constant}$ surface in the region $r(t) < r< r_\Lambda$, where $r_\Lambda$ is a radial cutoff with the interpretation of a fixed energy cutoff in the E-frame, i.e. $r_\Lambda = \Lambda$. This UV cutoff naturally restricts the time variable to $t< \pi/2 - O(\Lambda^{-1})$ on the approach of the domain wall to the boundary. The volume of the $\Sigma_{\rm UV}$ surfaces gives a universal UV contribution to the complexity given by
\eqn\uvc{
C_{\rm UV} (t) = {\Omega_{d-1} \over 4G} \int_{r(t)}^\Lambda {dr r^{d-1} \over \sqrt{1+r^2}}\;,
}
where $\Omega_{d-1}$ is the volume of the ${\bf S}^{d-1}$ sphere, giving the spatial volume $V_{\rm E}$ of the E-frame metric.  For $M\gg 1$ we can approximate $1+r^2 \approx r^2$ over the whole domain of integration and we have
\eqn\uvcmm{
C_{\rm UV}(t)_{M\gg 1} \approx {\Omega_{d-1} \over 4G(d-1)} \left(r_\Lambda^{d-1} - r(t)^{d-1}\right) \sim N_+ V_{\rm E}\,\left(\Lambda^{d-1} - M(t)^{d-1}\right)\;.
}
The result is physically sensible, being  proportional to $S_\Lambda - S_{M(t)}$, where $S_T$ is the high-temperature entropy at temperature $T$. It is always a decreasing function of  E-frame time, since the scale $M(t)$ separating UV from IR degrees of freedom is itself increasing.  For $M\ll 1$, we have a similar result for the purely UV complexity, except that the factor of $M(t)^{d-1}$ is replaced by $(\cos(t) \ell_+)^{1-d} = \cos(t)^{1-d}$. Explicitly,
\eqn\uvcmmm{
 C_{\rm UV}(t)_{M\ll 1} \approx {\Omega_{d-1} \over 4G(d-1)} \left(r_\Lambda^{d-1} - r(t)^{d-1}\right) \sim N_+ V_{\rm E}\,\left(\Lambda^{d-1} - \cos(t)^{1-d}\right)\;,
}
where we have also assumed that $t$ is large enough so that $r(t) \gg 1$. 

If the relevant operator gaps the system, or leaves out an infrared theory with few degrees of freedom, $C_{\rm UV}$ gives the complete answer for the complexity. On the other hand, if the bubble geometry has a smooth interior, there are infrared contributions.
The simplest answer is obtained for $M\gg 1$, with an interior AdS$_-$ bounded by the accelerating bubble wall. In this case a similar
calculation to \uvc\ yields
\eqn\irc{
C_{\rm IR} (t) = {\Omega_{d-1} \over 4G}\int_0^{r(t)} {dr r^{d-1} \over \sqrt{1+ r^2 / \ell_-^2}}\;,}
where we have included the effect of larger curvature in the interior AdS, i.e. $\ell_- < \ell_+ =1$. Since $r(0) =r_M \approx M \gg 1$, we can approximate the integral by 
$$
C_{\rm IR} (t) \approx {\Omega_{d-1} \ell_- \over 4G (d-1)} r(t)^{d-1}
$$
The presence of $\ell_-$ in place of $\ell_+ =1$ means that the overall factor of the IR complexity is slightly smaller than the UV one, consistent with the fact that the relevant operator does leave less degrees of freedom in the IR. Hence, we find that the total complexity {\rm decreases} as we approach the singularity in the E-frame. 

\bigskip
\centerline{\epsfxsize=0.7\hsize\epsfbox{figu2.eps}}
\noindent{\ninepoint\sl \baselineskip=2pt {\bf Figure 2:} {\ninerm
Extremal codimension-one surfaces  for dS-CFT crunch models with $M\gg 1$ (left) and $M\ll 1$ (right). }}
\bigskip

The case of a thick bubble wall, arising for $M \ll 1$, ultimately gives a similar answer. Here we emulate the situation which was originally discussed for black holes. The extremal surface in the interior of the bubble, $\Sigma_{\rm IR}$,  tends to lie along the maximal hyperbolic section at $\eta = \eta_m$ up to a hyperbolic radius of order $ \chi_{\rm exit} \sim \tau$. It  matches to the asymptotic surface $\Sigma_{\rm UV}$  through a small transition region, as shown schematically in figure 2. The volume of the IR contribution gives 
\eqn\irc{
C_{\rm IR} (t)     \approx {a_m^d \Omega_{d-1} \over 4G} \int_0^{\chi_{\rm exit}} d\chi \sinh^{d-1}  (\chi) \approx {a_m^{d} \Omega_{d-1}\over 4G} \,{e^{(d-1)\tau} \over 2^{d-1} (d-1) }
\;,}
a result that holds as $\tau \rightarrow \infty$, so that we can replace
$e^{\tau} /2 \approx 1/\cos(t) $ and find 
\eqn\finir{
C_{\rm IR} (t)_{M \ll 1} \approx N_+ V_{\rm E}\,a_m^d \cos(t)^{1-d}\;}
for the IR complexity on approach to the crunch singularity. 
Hence, the IR contribution has the same form as the second term in \uvcmmm, with opposite sign, the precise cancellation only occurring for the degenerate case of pure AdS geometry. 

 The behavior of the total complexity near the $M\ll 1$ crunches depends on whether   $a_m$ is larger or smaller than unity. 
We now give a somewhat more detailed argument 
to expect $a_m <1$. The FRW cosmology which describes the bubble interior must necessarily contain extra classical degrees of freedom, such as scalar fields $\Phi$, dual to the relevant operator $\CO$. The $M\ll 1$ background carries a small excitation of these fields around  the origin at $\rho=0$,  satisfying 
$$
\partial_\rho \Phi \Big |_{\rho =0} =0\;,
$$
for smoothness of the classical solution. When this solution is continued over the FRW patch across the horizon, the initial conditions on the scalar fields for $\eta$-evolution are $\Phi(\eta=0) \neq 0$ and $\partial_\eta \Phi(0) = 0$. Hence, the interior FRW cosmology starts with slightly displaced scalar fields, initially at rest in their potential. A crucial property is that, the $\CO$ operator being relevant, the dual scalar fields must be slightly tachyonic in AdS, albeit still satisfying the BF bound to preserve unitarity. Hence, the potential energy
for small fields starts with a negative quadratic term of order $\shalf m^2 \Phi^2$ with $-d^2 / 4 < m^2 <0$. This means that the energy density contributed by $\Phi$ on FRW hyperbolic slices 
$$
\rho_\Phi = \half (\pt_\eta \Phi)^2 + \half m^2 \Phi^2
$$
starts out {\it negative} at $\eta =0$. Since the FRW model is initially expanding, cosmological friction damps the kinetic energy of the scalar field as it rolls further down the inverted potential, implying that $
\rho_\Phi$ stays negative for some time beyond the turnaround time $\eta_m$.  The negativity of $\rho_\Phi (\eta_m)$ has consequences for the value of $a_m$ through the Friedman equation 
\eqn\few{
\left({da \over d\eta}\right)^2 + a^2 -{16\pi G \over d(d-1)} \rho_\Phi \,a^2 -1=0\;,
}
where we have used the normalization $\ell=1$ for the pure AdS solution $a(\eta)_{\rm AdS} = \sin(\eta)$. At the turning point $\eta_m$
the derivative vanishes and we obtain 
\eqn\caw{
a_m^2 = {1 \over 1-{16\pi G \over d(d-1)} \rho_{\Phi}(\eta_m)} \;,}
so that $\rho_\Phi (\eta_m) <0$ implies $a_m <1$.

\subsec{Complexity Of The Topological Crunch}

\noindent

The extremal surface for the topological crunch model wraps the ${\bf S}^1$ factor and the complexity should be normalized by the $d+2$ dimensional Newton's constant. In the AdS$_{d+1}$ directions, its structure is similar to that of the dS-crunch models.  There is a `thick bubble', whose extension outside the horizon is the region $0<\rho<1$ in the coordinates \bscc, defined by the fact that the radius of the compact circle remains approximately constant, equal to $R=1/2\pi M$. Therefore, this special region is seen as expanding along dS-invariant trajectories in the frame adapted to the E-frame. The interior of the horizon also has a  FRW patch, given by \frwc. Here, extremal surfaces tend to sit at some $\eta={\rm constant}$ surface which maximizes the $(d+1)$-dimensional volume factor
$
\cos(\eta)\,\sin^d (\eta)
$. This condition selects the critical space like surface at $\eta_m$ such that $(d+1) \cos^2 (\eta_m) =1$. The corresponding maximal value of the warp factor is
$$
f_m = \cos(\eta_m) \sin^d (\eta_m) = {d^{d\over 2} \over (d+1)^{d+1 \over 2}} <1\;.
$$
Since this quantity emulates the role of the critical scale factor $a(\eta_m)$ in the previous section, we see that this model satisfies the same inequality that was deduced in \caw. The contribution to the complexity coming from such surfaces cutoff at $\chi_{\rm exit} \sim \tau \rightarrow \infty$ is
$$
C_{\rm IR} \sim {1\over 4MG_{d+2}} f_m \int_0^\tau d\chi \sinh^{d-1} (\chi) \sim {1 \over 4MG_{d+2}} f_m {1 \over d-1} (e^\tau /2)^{d-1} 
$$
or, using a definition $MG_{d+2} = G_{d+1}$, 
\eqn\irt{
C_{\rm IR} \sim  {1\over 4G_{d+1}}{f_m \over d-1}  \cos^{1-d} (t)
\;.}
We see that it grows as $1/(t_\star -t)^{d-1}$ as we approach the crunch at $t_\star = \pi/2$. 

A technical  difficulty of the topological crunch model is the complicated structure of the extremal surfaces in the UV. In the asymptotic region $\rho\gg 1$ these surfaces are no longer lying at $t={\rm constant}$ sections, nor do they correspond to $\tau={\rm constant}$ sections. A lower bound on the complexity can be given by evaluating it along $t={\rm constant}$ sections, which gives the result
\eqn\uvt{
C_{\rm UV} \geq {1\over 4G_{d+1}} \int_{r(t)}^\Lambda dr\,r^{d-1} \,\cos(t) \sim {1\over 4G_{d+1}} {1\over d} \cos(t)\, \left(\Lambda^d - {1\over \cos(t)^d}\right)
\;,}
where we have used $r(t) \sim 1/\cos(t)$ at large $t$.  The contribution of the lower endpoint is proportional to $-\cos(t)^{1-d}$, having the same time dependence but opposite sign from the IR contribution found in \irt. If we compare the two coefficients in front we
get
$$
{f_m\over d-1}  - {1\over d}
$$
as controlling the overall sign of the diverging $\cos(t)^{1-d}$ term.  This quantity is always negative for  $d>1$. So we again find that  the diverging term decreases. The term which depends on the UV cutoff $\Lambda$ is positive but also decreasing because of the volume of the shrinking circle. So we conclude that this estimate still has decreasing complexity towards the singularity. Although in this case this is based on a set of test surfaces with sub-optimal volume, we expect this result to capture the qualitative behavior of the true complexity functional, since the biggest difference between the true extremal surfaces and the $t={\rm constant}$ ones occurs near the boundary, but we know that the UV contribution to the complexity must be small due to the shrinking volume of the CFT frame.

\subsec{Complexity Of The Kasner-AdS Crunch}
\noindent

We start by rewriting the bulk metric of the Kasner model \kb\ in the form
\eqn\kz{
ds^2 = {-dt^2 + \sum_i t^{2p_i} dx_i^2 + dz^2 \over z^2}
\;,}
by the change $r=1/z$. 
In computing the complexity of this state,  we let the codimension-one surface extend in the $x_i$ directions and be described by a function $t(z)$ as it enters the bulk from the $z=0$ boundary. The volume functional is then given by (for $t(z)>0$) 
\eqn\volf{
{\rm Vol} (\Sigma) = V_x \int {dz \over z^{d}} t(z) \,\sqrt{1-t'(z)^2}\;,
}
where $t'(z) = dt /dz$ and $V_x$ is the comoving volume in the  $x_i$ coordinates.  Extremal surfaces satisfy the Euler--Lagrange equation
\eqn\ela{
{d \over dz} \left[{-t(z)t'(z) \over z^d \sqrt{1-t'(z)^2}}\right] = {\sqrt{1-t'(z)^2} \over z^d}\;,
}
Operating and removing the square roots we obtain the related equation
\eqn\redu{
z t t'' + z(1-t'^2)t'^2 -d \cdot tt'(1-t'^2) + z(1-t'^2)^2 =0\;.
}
Numerical analysis shows that the solutions of this equation with boundary conditions $t(0)=t, t'(0) = t'_0$ are asymptotic to $t(z) \sim \pm z $ as $z\rightarrow \infty$. We can confirm this by writing the ansatz
$$
t(z) = \pm z + \varepsilon(z)
$$
and keeping only terms linear in $\varepsilon (z)$ and its derivatives, we find an equation which only determines $\varepsilon'(z)$, i.e.
$$
z \varepsilon'' +2(d-1) \varepsilon' =0\;,
$$
which ensures the null asymptotics as $z\rightarrow \infty$ with small corrections of order $z^{2-2d}$. Further evidence in favor of the asymptotically null character of the solutions comes from the existence of an exact solution of \redu\ given by a straight-line $t(z) = \pm z$. While this is not a proper solution of the original equation \ela, it shows that solutions with null asymptotics should exist. 

The asymptotically null character of $\Sigma$ at large $z$ means that the volume \volf\  is dominated by the UV contribution. Near the $z=0$ boundary, we can evaluate the volume contained between two cutoff slices $z_\Lambda = 1/\Lambda$ and $z_{2\Lambda} = 1/2\Lambda$, in the linearized approximation where $t(z) \approx t + t'_0 z$. We find
$$
{\rm Vol}(\Sigma)_{\Lambda}^{2\Lambda} = V_x \int_{z_{2\Lambda}}^{z_\Lambda}  {dz \over z^d} (t + t'_0 z)\sqrt{1-(t'_0)^2} \propto V_x \Lambda^{d-1} (t + a \,t'_0 /\Lambda) \sqrt{1-(t'_0)^2}
$$
where $a$ is an $O(1)$ positive constant. In the limit in which we decouple the cutoff, $\Lambda \rightarrow \infty$, we find that the
term proportional to $a$ is irrelevant, and we remain just with the term proportional to $t \sqrt{1-(t'_0)^2}$, which is maximized by
$t'_0 =0$. Hence the UV contribution to the volume functional is of order
$$
{\rm Vol}(\Sigma_t) \sim V_x \,\Lambda^{d-1} \,|t|
$$
and the corresponding complexity scales as
\eqn\complekasner{
C(t) \sim N^2 V_x\, \Lambda^{d-1} \,|t| \sim N^2 V_{\rm CFT} \,\Lambda^{d-1}\;.
}
The outstanding consequence is the formal vanishing of the complexity at the Kasner singularity. More precisely, the complexity scales extensively with the physical volume of the CFT frame. Since $\Lambda^{-1}$ is a physical `lattice cutoff', it is not reasonable to follow the cosmological evolution of a given comoving domain down to times such that $V_{\rm CFT} \Lambda^{d-1} <1$. This introduces a characteristic UV cutoff to the time variable as we approach the singularity, depending on the extension of comoving volume we are considering. 

\newsec{Discussion}

\noindent

We have analyzed the behavior of the volume/complexity proposal for certain AdS/CFT realizations of singular cosmologies. Unlike the singularities cloaked inside black hole horizons, these models harbor singularities which are visible by any freely-falling observer in the bulk space-time. In the language of the AdS/CFT correspondence, one can say that these singularities can be efficiently probed by local operators 
in the dual field theory. We have performed estimates in three classes of scenarios and found a tendency for the so-defined holographic complexity to {\it decrease} as the singularity is approached. Taken at face value, this means that the quantum state has a simpler entanglement structure when close to the singularity. 

The mechanism for this phenomenon is interesting, since complexity is identified with a certain maximal-volume space-like slice,  
standard crunch singularities tend to `repel' the extremal surface. A time slicing which gets close to the singularity must probably require an extremely non-local Hamiltonian in the dual QFT picture.  In models were the complexity is dominated by the UV, like the Kasner crunch model, the small volume is instead achieved by the  extremal surface becoming approximately null in the IR. In models where the IR contribution does grow with time, with a behavior similar to that of eternal black holes,  it turns out that the IR/UV interphase  is time-dependent, and the overall balance favors the UV by standard properties of the renormalization group.

This result contrasts to some extent with the intuition gained by the propagation of probes near crunch singularities, where the high energy blueshift causes unbounded local excitation: singularities are regarded as `hot' and full of complicated dynamics. A generic crunch is usually imagined as a complicated pattern of colliding black holes. On the other hand, the classic BKL work \refs\bkl\ shows that there are universal features to singularities at the ultra-local level (see also the modern work \refs\damour). It is hard to asses the {\it a priori} relevance of this fact, since it depends to a large extent on the particular dynamics of Einstein gravity, but it is tempting to adventure a possible relation between the BKL universality and the low holographic complexity found here.

\bigskip{\bf Acknowledgements:} 

We thank E. Rabinovici for discussions. This work   was partially supported by MINECO and FEDER under a grant  FPA2012-32828, and the 
spanish MINECO {\it Centro de Excelencia Severo Ochoa Program} under grant SEV-2012-0249.

{\ninerm{
\listrefs
}}

\end

\newsec{dS-CFT Crunches}

\noindent

Consider a CFT defined on global $d$-dimensional de Sitter space-time (dS):
\eqn\ds{
ds^2_{{\rm dS}_d} = -d\tau^2 + \cosh(\tau)^2 \,d\Omega_{d-1}^2\;,
}
where we work in units of the Hubble parameter. Bulk duals preserving the dS symmetries have the form
\eqn\sds{
ds^2_{\rm bulk}  = d\rho^2 + f(\rho)^2 \,ds^2_{{\rm dS}_d}\,
}
with the function $f(\rho)$ determined by Einstein's equations in the bulk, with negative cosmological constant. The standard vacuum solution  is  $f(\rho)= \sinh(\rho)$, giving the `dS wedge' of AdS,
\eqn\dspatch{
ds^2_{\rm dS\; wedge} = d\rho^2 + \sinh^2 (\rho) \left(-d\tau^2 + \cosh(\tau)^2 \,d\Omega_{d-1}^2 \right)\;. 
}
This metric covers a proper subset of global AdS, defined by all the points sitting at space-like distance from the  $t=r=0$ `center' of AdS in the global parametrization
\eqn\globalp{
ds^2_{\rm global} = -(1+r^2) dt^2 + {dr^2 \over 1+r^2} + r^2 \,d\Omega_{d-1}^2\;.
}
The explicit change of variables between \dspatch\ and \globalp\ is given by
\eqn\chang{
r = \sinh(\rho) \cosh(\tau)\;, \qquad \cos (t) = {\cosh(\rho) \over \sqrt{1+\sinh^2 (\rho) \cosh^2 (\tau)}}\;,
}

\subsec{An Exact Crunch Model}

\noindent

There is an  interesting  exact realization of \buint\  which is obtained starting with a $d+2$ dimensional Euclidean AdS space
in the global parametrization
\eqn\glof{
ds^2_{d+2} = \cosh^2 (\rho)\, d\phi^2 + d\rho^2 + \sinh^2 (\rho) \,d\Omega_d^2
\;,}
and defining the particular Lorentzian continuation in which the $d$-sphere turns into dS$_d$, namely
\eqn\glol{
ds^2_{d+2} = \cosh^2 (\rho) d\phi^2 + d\rho^2 + \sinh^2 (\rho) (-d\tau^2 + \cosh^2 (\tau) d\Omega_{d-1}^2 )\;.
}
If we identify the $\phi$ coordinate with period $M^{-1}$, the conformal boundary of this metric is ${\bf S}^1 \times {\rm dS}_d$  where the circle has size $M^{-1}$. Hence, placing a CFT on this space-time induces a massive theory on dS$_d$ with effective mass $M$. In the bulk, the `core region' $0< \rho \ll 1$ is approximately the product of a fixed size ${\bf S}^1$ times the `dS-wedge' metric \dspatch.  Continuing now across the $\rho=0$ horizon by writing 
$
\rho = i \eta$ and  $
\tau = \chi - i\pi/2
$, we end up with 
\eqn\intfd{
ds^2 = -d\eta^2 + \sin^2 (\eta) \left(d\chi^2 + \sinh^2 (\chi) d\Omega_{d-1}^2 \right) + \cos^2 (\eta) \,d\phi^2 
\;,}
which consists on a circle of size $\cos(\eta)/M$ fibered over the FRW patch of vacuum AdS$_{d+1}$. The complete $d+2$ dimensional geometry is singular at $\eta_\star = \pi/2$, corresponding to the point where the compact circle shrinks to zero size. Alternatively, if we were to `integrate out' the circle, we would obtain a singular model in the \buint\ class. 

we specify maximal codimension-one surfaces   with the property that they approach a constant-time slice at the AdS boundary. This asymptotic condition depends on the particular frame of interest, namely for dS models we have
either $\tau={\rm constant}$ asymptotics on the dS-frame or $t={\rm constant}$ asymptotics in the E-frame. In each case, the asymptotic surface $\Sigma_{UV}$ makes a contribution of order $N^2 V \Lambda^{d-1}$ to the complexity, where $V$ is a volume in the corresponding CFT metric and $N^2$ is the number of degrees of freedom, of the order of the CFT's central charge. $\Lambda$ is an energy UV cutoff adapted to the particular frame of interest.  We illustrate in Figure 2 the different geometry of the asymptotic extremal surfaces in the two standard frames, for the particular case of the vacuum AdS manifold.

\bigskip
\centerline{\epsfxsize=0.7\hsize\epsfbox{figu2.eps}}
\noindent{\ninepoint\sl \baselineskip=2pt {\bf Figure 2:} {\ninerm
Extremal codimension-one surfaces in the dS-frame (left) and E-frame (right). Also shown are the dS-invariant (left) and E-invariant (right) UV cutoff surfaces.  The volume of each cutoff surface is proportional to $\Lambda_{\rm dS}^{d-1}$ (left) and $\Lambda_{\rm E}^{d-1}$ (right).}}
\bigskip

In the case of the perturbed states, there is an asymptotic contribution of order 
$$
{{\rm Vol}(\Sigma_{UV}) \over 4G}  \sim N^2 V_{\rm dS} (\Lambda_{\rm dS}^{d-1} - M^{d-1})
$$
 in the dS frame, and one of order 
 $$
 {{\rm Vol}(\Sigma_{UV}) \over 4G}  \sim N^2 V_{\rm E} (\Lambda_{\rm E}^{d-1} - M(t)^{d-1})
 $$
  in the E frame.   These UV contributions are the same for all states, including the gapped states whose holographic dual is given by an expanding bubble of nothing, a case where it gives the complete answer. In the cases where there is a bubble with an interior, the complexity surface is continued through the interior as a maximal volume surface $\Sigma_{IR}$,  with a Dirichlet boundary condition at the scale of the bubble wall. The simplest construction occurs for the limit of a thin wall, i.e. we just have $t= {\rm constant}$ surfaces of the interior E-frame, matched at the domain wall to the asymptotic portion. We show in Figure 3 the corresponding surfaces in the two frames of interest.  

\bigskip
\centerline{\epsfxsize=0.7\hsize\epsfbox{figu3.eps}}
\noindent{\ninepoint\sl \baselineskip=2pt {\bf Figure 3:} {\ninerm
Extremal surfaces in the thin-wall approximation, for the dS frame (left) and E frame (right).}}
\bigskip

The interior geometry  appearing in Figure 3 has a  smaller radius of curvature, corresponding to the smaller number of degrees of freedom left below the  scale $M$, i.e. $N^2_{IR} < N^2$. The total E-frame complexity reads 
\eqn\efc{
C_{\rm E} (t) \approx N_{IR}^2 V_{\rm E} M(t)^{d-1} + N^2 V_{\rm E} (\Lambda_{\rm E}^{d-1} - M(t)^{d-1}) \;,
}
where the time-dependent mass scale diverges as $M(t) \sim M \Omega(t) \sim M/(t_\star - t)$. Since $N_{IR} < N$, the see that the complexity actually {\it decreases} as $t$ approaches the singularity at $t_\star = \pi/2$. 

For the dS-frame case, we have instead
\eqn\dsfc{
C_{\rm dS} (\tau) \approx N_{IR}^2 V_{\rm E} M(t)^{d-1} + N^2 V_{\rm dS} (\Lambda_{\rm dS}^{d-1} - M^{d-1})
\;.}
Since $V_{\rm E} M(t)^{d-1} = V_{\rm E} \Omega(t)^{d-1} M^{d-1} = V_{\rm dS} M^{d-1}$, we find that the dS-frame time dependence of the complexity comes only through the expanding volume of dS space.

\bigskip
\centerline{\epsfxsize=0.7\hsize\epsfbox{figu4.eps}}
\noindent{\ninepoint\sl \baselineskip=2pt {\bf Figure 4:} {\ninerm
Extremal surfaces in the $M\ll 1$ limit, for the dS frame (left) and E frame (right).}}
\bigskip

The case of a thick bubble wall, arising for $M \ll 1$, ultimately gives a similar answer. Here we emulate the situation which was originally discussed for black holes. The extremal surface in the interior of the bubble, $\Sigma_{IR}$,  tends to lie along the maximal hyperbolic section at $\eta = \eta_m$ up to a hyperbolic radius of order $ \chi_{\rm exit} \sim \tau$. It  matches to the asymptotic surface $\Sigma_{UV}$  through a small transition region. The volume of the IR contribution is
\eqn\irc{
{{\rm Vol} (\Sigma_{IR}) \over 4G}   \sim {f_m^d \over 4G} \int_0^{\chi_{\rm exit}} d\chi \sinh^{d-1}  (\chi) \sim {f_m^{d} \over 4G} \,{e^{(d-1)\tau} \over (d-1) }
\;,}
which scales like the CFT volume in dS frame, $V_{\rm dS} \sim e^{(d-1)\tau}$, just as the UV contribution. 

Since the warp function $f(\eta)$ is approximately equal to the vacuum warp function $\sin (\eta)$ for $\eta \approx 0$, but it vanishes
earlier, at $\eta_\star < \pi$, it is natural to suppose that $|f(\eta)| < |\sin(\eta)|$ for $0<\eta<\eta_\star$. Under this assumption, ${\rm Vol}(\Sigma_{IR})$ is strictly smaller than the volume of the same surface in the vacuum metric \vfrw. In turn, the hyperbolic surface in the vacuum metric anchored at $\chi_{\rm exit} \sim \tau$ has smaller volume than the E-frame maximal volume metric exiting at the same point. For the
vacuum metric the total volume of $\Sigma_{IR} \cup \Sigma_{UV}$ remains constant in time. Therefore we conclude that the E-frame extremal surface in the warped geometry has {\it decreasing} volume and thus we again have a decreasing complexity in the frame which {\it sees} the singularity. 

We have seen that those singularities that are engineered by singular driving of relevant operators have a `low' complexity in those frames that meet the singularity `head on'. By low complexity we mean that it remains bounded and even decreases in the time variable which parametrizes the distance to the singularity.

To be more specific, let us consider a standard large eternal black hole in AdS$_{d+1}$ with spherical horizon. The exterior metric corresponds to the $f(r) >0$ region of 
\eqn\etbh{
ds^2 = -f(r) dt^2 + {dr^2 \over f(r)} + r^2 d\Omega_{d-1}^2\;,
\qquad f(r) = 1+ {r^2 \over \ell^2} - {\mu \over r^{d-2}}\;.}
Large AdS black holes are characterized by a horizon radius $r_0 \gg \ell$, and satisfy the scaling $\mu \sim r_0^d / \ell^2 \sim \ell^{2d-2} T^d $. The shared interior can be regarded as the $0<r<r_0$ region, where $r$ is now a time-like coordinate and the metric takes the form 
\eqn\met{
ds^2 = |f(r)|\,d{\bar t}^{\,2} - {dr^2 \over |f(r)|} + r^2 \,d\Omega_{d-1}\;.
}
The extremal interior surface, denoted $\bar\Sigma$,  sits at  the particular value of $r$, denoted $r_m$,  which maximizes the volume element $ |\Delta {\bar t} \,|\, r^{d-1} \sqrt{|f(r)|} 
$
 for a fixed interval of the space-like ${\bar t}$ coordinate. 
 For any finite  but large value of $t$, the extremal surface $\Sigma_t$ is well-approximated by the portion $\Delta {\bar t} \approx t$ of this $r_m = {\rm constant}$ surface in the interior,  joined through a transition region of $O(r_0)$ length  to the $t={\rm constant}$ surface in the exterior.  In  regular black holes, $r_m$ is parametrically of order $r_0$, and the extremal surface $\Sigma_t$  is smooth, staying clear from the singularity at $r=0$ for all values of $t$. Therefore, for $r_0 \gg \ell$ its interior portion makes a contribution   to \ansatz\ of order $r_m^{d-1} \,\sqrt{|f(r_m)|} /G \ell \sim S \,T\,t$. 
 
 The exterior portion of $\Sigma_t$ has infinite volume, and we may cut it off at large radius, giving a purely UV contribution to the complexity scaling like
 \eqn\cuv{
 C_{\Lambda} \sim V \,{r_\Lambda^{d-1} \over G} \sim N^2 \,V \,\Lambda^{d-1}\;,
 }
  where $N^2 = \ell^{d-1} /G$ gives the effective number of CFT degrees of freedom and $V\Lambda^{d-1}$ is the CFT spatial volume in units of the UV lattice spacing $\Lambda^{-1} = \ell^2 / r_\Lambda$. Thus, we obtain the  estimate $C(t) \sim C_\Lambda + S T t$ at very large $t$, where the UV contribution $C_\Lambda$ is time independent. Along the way, we also learn that $\Sigma_t$ is actually a  {\it maximal}  codimension one surface.

 The saturation of the $\bar\Sigma$ surface  at  a fixed `distance' from the singularity prevents the $C(t)$ observable from `proving' it, at least as far as we remain within the leading large-$N$ approximation. Since the ansatz \ansatz\ singles out maximal surfaces, this property of `singularity avoidance' seems to be rather generic for any kind of  `bang' or  `crunch'. We learn that, despite their special status at the cosmological `ends of time', bangs and crunches in the interior of black holes make a comparatively small  contribution to the long-time computational complexity of those quantum states. 
 
 In this note we seek to put this notion of singularities as `computationally simple' objects on a firmer basis by examining a number of examples in which the holographic interpretation of the singularities becomes sharper. In particular, black hole singularities are quintessential `interior' objects, and thus are hidden in the deep subtleties of infrared observables of the CFT. Fortunately, there are simple constructions of holographic states with space like singularities which are  visible in the UV sector of the CFT. The bulk geometries look like crunchy versions of AdS cosmologies and the dual CFT description involves a singular driving of the CFT Hamiltonian by terms of general  form
\eqn\drives{
H(t)= H_{\rm CFT} + \int_{\bf X} J(t) {\cal O}
\;,
}
with ${\cal O}$ a relevant or marginal operator and $J(t)$ a function with prescribed singular behavior at some finite time $t=t_\star$. More concretely, we will examine two broad classes of models: crunch singularities induced by relevant perturbations of CFTs on de Sitter space, and CFTs on Kasner geometries as an example of singular driving by an exactly marginal operator. 
 In all cases, we will see that the holographic definition of complexity \ansatz\ assigns  low values to such singular states, as seen in a conformal frame which meets the singularity in finite time.

\newsec{Singularities, Conformal Frames And Complexity}

\noindent 

Singular states with AdS/CFT interpretation can be defined by singular driving, according to the scheme in \drives. The mildest version uses relevant operators ${\cal O}$, with conformal weight $\Delta < d$, so that we can write for the source 
$$
J(t) = M(t)^{d-\Delta} 
$$
where $M(t)$ is a time-dependent mass scale which diverges at some finite time $t_\star$. If such a driving is specified in a particular static conformal frame $X_d = {\bf R} \times {\bf X}_{d-1}$  with CFT metric
$$
ds^2_d = -dt^2 + ds^2_{\bf X} 
$$
a time reparametrization $t\rightarrow \tau (t)$ which postpones the singularity to $\tau=\infty$ and defined by some rescaling function $\Omega(t)$, 
$$
{d\tau \over dt} = \Omega(t)
$$
allows us to write
$$
ds^2 = \Omega(t)^{-2} \left[ -d\tau^2 + \Omega(\tau)^{2} ds^2_{\bf X}\right] 
$$
The metric in brackets defines a new conformal frame ${\widetilde X}_d$ of FRW type, in which the driving term becomes
$$
{\widetilde H} = H_{\rm CFT} + \int_{\widetilde {\bf X}} {\widetilde M}(\tau)^{d-\Delta}\,{\widetilde \CO} 
$$
with ${\widetilde M}(\tau) = M(t) / \Omega(t)$. 
If we now chose $\Omega(t)$ to cancel the time dependence of $M(t)$, we could have started our discussion by presenting a relevant, static deformation of a CFT on the FRW manifold, with no obvious singularity in sight, and recast the dynamics as a singular driving on 
a static frame. The simplest such example uses global de Sitter space-time, dS$_d$,  as the FRW frame and the Einstein universe ${\rm E}_d = {\bf R} \times {\bf S}^{d-1}$ as the static frame. 

The CFT in the FRW frame has a dimensionless parameter, given by the ratio between the Hubble scale and the $\tau$-independent  mass scale ${\widetilde M} = M(t)/ \Omega(t)$, equivalently the product ${\widetilde M} \ell$, where $\ell$ is the dS$_d$ curvature radius.

\end

There are also tractable examples where the CFT is defined on a static and smooth space-time:  
$$
ds^2_{\rm CFT} = -dt^2 + d\ell^2
$$
with $d\ell^2$ the metric of some purely spatial manifold. In this situation, the singular driving may be achieved by connecting a relevant operator $\CO$ with weight $\Delta < d$, 
\eqn\sd{
\delta \CL = \int_{X_d} M(t)^{d-\Delta} \CO\;, }
where $M(t)$ is a time-dependent mass scale which becomes infinite at some finite $t=t_\star$. It is interesting to notice that any model of this type can be presented in a frame which hides the singularity at `future infinity' in the appropriate time variable. To this end, we define a reparametrization $t \rightarrow \tau(t)$ with the property that $\tau(t_\star) = \infty$. Defining 
$$
{d{\tau} \over dt} = \Omega(t)
$$
the original frame arises then as a Weyl rescaling of a FRW frame ${\widetilde X}_d$, 
$$
ds^2 = \Omega(t)^{-2} \,{\widetilde {ds}}^{\,2}
$$
with metric 
$$
{\widetilde {d\,s}}^{\,2} =-d{\tau}^{\,2} + \Omega({\tau}\,)^{2} d\ell^2 \;. 
$$
Written in the FRW frame, the driving term pics the effect of the conformal transformation 
$$
{\widetilde {\delta \CL}} = \int_{\widetilde  X_d} {\widetilde M}({\tau}\,)^{d-\Delta}\,{\widetilde \CO} 
$$
with ${\widetilde M}({\tau}\,) = M(t) / \Omega(t)$. 
If we now chose $\Omega(t)$ to cancel the time dependence of $M(t)$, the scale ${\widetilde M}(\tau) =M$ will be $\tau$-independent. We could have started our discussion by presenting a relevant, $\tau$-invariant deformation of a CFT on the FRW manifold, with no obvious singularity in sight, and later recast the dynamics as a singular driving on 
a $t$-invariant  frame. The simplest such example uses global de Sitter space-time, dS$_d$,  as the FRW frame,
\eqn\dsframe{
ds^2_{\rm dS} = -d\tau^2 + \cosh^2 (\tau) \,d\Omega_{d-1}^2\;,
}
 and the Einstein universe ${\rm E}_d = {\bf R} \times {\bf S}^{d-1}$,
 \eqn\eframe{
 ds^2_{\rm E} = -dt^2 + d\Omega_{d-1}^2\;,}
 as the static frame. Both frames are 
 related by a conformal map with $\Omega(\tau) = \cosh(\tau) = 1/\cos(t)$, thus becoming singular at $t_\star = \pm \pi/2$.  In what follows, we will focus on such a model, whose dS-frame CFT is perturbed by a de Sitter-invariant mass scale $ M$, and its E-frame counterpart has a diverging mass scale $M(t) = M \Omega(t)$.  A classical picture which serves to shape intuition is provided by a Landau--Ginzburg scalar potential in $d>2$ dimensions:
\eqn\lg{
V(\Phi) = \pm M^2 |\Phi|^2 + \lambda |\Phi|^{2d \over d-2}
}
with $\lambda >0$ and $ M^2 \gg 1$. This last condition means that the effects of the mass deformation parameter $M$ will dominate over the finite-size effects of either the dS frame (Hubble scale) or E frame (size of the spatial sphere). 
 A   positive sign in \lg\  gives   a mass gap of order $M$, whereas a negative sign induces symmetry breaking and  possibly an infrared CFT. 
 
 Although the strongly coupled CFTs with AdS duals are far from the simple classical model in \lg, one can recognize the qualitative dynamical features in the bulk AdS dynamics: a de Sitter-invariant `bubble of nothing' in the first case and  a de Sitter-invariant  domain wall in the second case.         In the Einstein-universe frame, these de Sitter-invariant defects are seen to hit the AdS boundary in finite time. The case of the domain wall bounding a bubble with nontrivial `interior'  is especially interesting, since the finite-time hit carries an infinite amount of kinetic energy into the boundary, producing a singularity which tears AdS from the boundary impact point into the interior.  

In the case that the relevant operator at $M\gg 1$ leaves a fully fledged large-$N$ CFT in the infrared, it makes sense to 
 approximate the bulk picture by a thin-wall interphase between two AdS spaces of different curvature. \foot{ The interior AdS is more curved, revealing the fact that the IR CFT has less degrees of freedom than the UV CFT.} In this case the interior is largely given by a vacuum AdS metric, and the singularity develops `late', being almost null. A good parametrization is given by the FRW patch of AdS,
\eqn\vfrw{
ds^2 \approx -d\eta^2 + \sin^2 (\eta) \left(d\chi^2 + \sinh^2 (\chi) \,d\Omega_{d-1}^2 \right)
\;,}
with the singularity developing closer to the null surface  $\eta=\pi$ the better is the thin-wall approximation. 

In actual CFT situations one must give some concrete UV regularization of $C(t)$. Since we are interested on singularities which can be detected by UV probes, some care must be exercised with regularization prescriptions.   A simple procedure involves a radial cutoff $r\leq r_\Lambda$ in \asme, equivalent to a short distance cutoff $\Lambda^{-1} = \ell^2 / r_\Lambda$. This regularization is adapted to the particular conformal frame of choice, since it  preserves the isometries of the  CFT manifold $X_d = {\bf R} \times {\bf X}_{d-1}$.  The asymptotic region \asme\ gives a generic UV contribution to the complexity of the form    
 \eqn\cuv{
 C_{\Lambda} \sim V \,{r_\Lambda^{d-1} \over G} \sim N^2 \,V \,\Lambda^{d-1}\;,
 }
 where $V$ is the `spatial volume' along the ${\bf X}_{d-1}$ manifold and we have used the standard rule in AdS/CFT, relating the number of microscopic degrees of freedom $N^2$ with Newton's constant and the AdS curvature radius, $N^2 = \ell^{d-1} / G$. The answer \cuv\ scales like the microscopic entropy of states with coarse graining `temperature' of order $\Lambda$. This is perfectly compatible with the tensor network interpretation since vacuum CFT states are associated to tensor networks constructed over expander graphs, i.e. with an inherent hyperbolic connectivity, ensuring that volumes scale asymptotically as `areas', thus explaining the $\Lambda^{d-1}$ behavior of \cuv.    
 
 In  a sort of `Wilsonian interpretation', we can define the amount of complexity contributed by an energy band $\mu < E < \Lambda$ as the volume between radii $r_\mu$ and $r_\Lambda$:
 $$
 C_\Lambda - C_\mu \sim N^2 V (\Lambda^{d-1} - \mu^{d-1})
 $$
 This is a natural measure of complexity for a gapped state, with mass gap of order $\mu$.

\end

There are quite a few cases in quantum field theory where the same set of data can be described in several ways. The two dimensional Ising model and electric magnetic duality are examples. In string theory a very large set of such relations was uncovered over the years.
Backgrounds which have for example different metric, topology, number of small or large dimensions and singularity, commutativity and associativity structures were identified. The AdS/CFT type relations are in this class.  In an effort to come to grips with the special challenges presented by black hole physics a concept named Black Hole Complementarity was put forward \refs\comp. Sets of observables defined outside and inside the horizon,  while not commuting among the sets, were each supposed to give a full description of the system.
The consequences of such a suggestion are still being processed (cf. \refs\fire\ and its wake). 

In a previous paper we have been brought to suggest a relation which touches all these types of dualities \refs\usd.
The systems discussed, a certain type of crunching AdS space-times, have a cosmological horizon separating those observers who meet the crunch in finite proper time from those who get to live for an infinite proper time.  The situation is thus similar to a black hole of infinite entropy. It was claimed  in \refs\usd\ that the exterior physics can be described, via  AdS/CFT  tools, by a specific class of non-singular time-independent QFTs living on a time-dependent de Sitter (dS) world volume, whereas the horizon interior could be described by a time-dependent QFT living on a static Einstein universe. The two holographic descriptions are related by a conformal transformation, which becomes equivalent to a complementarity map for this system. 

The conformal complementarity relates the `eternal' Hamiltonian evolution of dS space-time to a finite-time interval of the Einstein universe, which we call `apocalyptic'. This property is visible in the short-distance description of the QFT, and can be studied with effective Lagrangian methods, something we address in sections 2 and  3. Furthermore, if the system is simplified by `dimensional reduction' to the conformal quantum mechanics of de Alfaro, Fubini and Furlan \refs\aff, the complementarity transformation becomes explicitly  expressible for any wave function in the Hilbert space.   We analyze in sections 4 and 5 the details of this $d=1$ system, including the mapping of observables on both sides of the duality. Section 6 is devoted to a formal extension of the eternal/apocalyptic duality to arbitrary QFTs and we end with a discussion of conceptual puzzles and open questions in section 7, where we also succumb to the temptation to relate these ideas to some features of our universe.

\newsec{A Simple Model Of Cosmological Complementarity}

\noindent

A relation between horizon complementarity and conformal symmetry is inherent in AdS/CFT as a result of
basic rules of the correspondence. An AdS$_{d+1}$ space-time does not define a canonical metric on the $d$-dimensional boundary but rather defines  a boundary conformal structure, i.e. a conformal class of $d$-metrics. Conformal maps between these metrics extend naturally as bulk  diffeomorphisms, whose global  properties  produce  some degree of ambiguity in the precise rules by which a given abstractly defined  CFT codifies the bulk geometry.  

To appreciate the point, let us consider the AdS$_{d+1}$ global manifold with metric
\eqn\eframe{
ds_{\rm global}^2 = -(1+r^2)\, dt^2 + {dr^2 \over 1+r^2} + r^2 \,d\Omega_{d-1}^2\;,
}
where lengths are measured in units of the AdS radius of curvature. 
According to the standard AdS/CFT rules 
\refs\adscft, we may regard \eframe\ as the vacuum state of a dual CFT on the Einstein manifold ${\rm E}_d={
\bf R} 
\times {
\bf S}^{d-1}$ with metric 
\eqn\emet{
ds^2_{{\rm E}} = -dt^2 + d\Omega_{d-1}^2\;.
}
Small perturbations of \eframe\ quantized on a low-energy effective field theory approximation can be regarded as low-lying excitations of the CFT on \emet\ and can be described by  a Hamiltonian picture for all values of the Einstein-frame time variable $t
\in {
\bf R}$.

Alternatively, we could have started with a different presentation of the  AdS$_{d+1}$ space-time, with a metric which we denote `the bubble':
\eqn\dsframe{
ds_{\rm bubble}^2 = d\rho^2 + \sinh^2 (\rho) \left(-d\tau^2 + \cosh^2( \tau) \,d\Omega^2_{d-1} \right),}
made out of a de Sitter foliation of AdS. 
Taking the $\rho \rightarrow \infty $ limit and rescaling by the divergent factor of $\sinh^2 (\rho)$ we have a different conformal boundary metric: 
\eqn\dsm{
ds^2_{\rm dS} = -d\tau^2 + \cosh^2 (\tau)\,d\Omega_{d-1}^2
\;,}
given by the global de Sitter manifold. Thus, we can also regard the version of AdS given by \dsframe\ as the bulk dual of the CFT on the dS$_d$ global manifold with metric \dsm\ (cf. \refs\mogollon). Not surprisingly, the two boundary metrics are conformally related by a Weyl rescaling and a time diffeomorphism:
\eqn\conf{
ds^2_{{\rm dS}} = \Omega^2 (\tau) \,ds^2_{{\rm E}
} 
\;, \qquad \Omega (\tau) = \cosh (\tau) = {1\over \cos (t)}\;,}
a map that should be  unitarily represented in the Hilbert space and operator algebra of the abstractly defined CFT.

 \bigskip
\centerline{\epsfxsize=0.2\hsize\epsfbox{rgr.eps}}
\noindent{\ninepoint\sl \baselineskip=2pt {\bf Figure 1:} {\ninerm
Radial slices adapted to E$_d$ and dS$_d$ isometries, corresponding to fixed $r$ and fixed $\rho$ respectively. Each point is a ${\bf S}^{d-1}$ sphere with radius ranging from zero at the origin of polar coordinates (left dashed line) to infinite size at the AdS boundary (right boundary line). }}
\bigskip

On the other hand,  the `bubble' version of AdS given by \dsframe, with coordinate domains  $-\infty < \tau<\infty$ and $0\leq \rho < \infty$, only covers a proper subset of the whole global AdS manifold \eframe: while the $r$-slices generate the whole AdS bulk, the $\rho$ slices only cover the causal diamond subtended by the $t\in  [-{\pi \over 2}, {\pi \over 2}]$ interval of E-time and bounded by the null surfaces $\rho=0$, $\tau = \pm \infty$.
 This raises the question of how the two CFT descriptions can be unitarily equivalent while one of the bulk duals is strictly contained into the other.

It turns out that the two bulk formulations are truly equivalent, in the sense that each one of them contains all the information needed to reconstruct the other \refs{\banksdo, \susfrei, \insightfull}. The key  fact making this equivalence possible is the existence of a common initial value surface in both bulk domains.
As shown in Figure 2, the Hamiltonian development of the dS-foliated patch shares an initial-value surface with the
E-foliation of the global AdS manifold, namely the $\tau=t=0$ surface. Therefore, any {\it perturbative} bulk state  defined on an arbitrary $t={\rm constant}$ surface may be unitarily  `pulled-back' to the $t=0$ initial-value surface, which coincides with the $\tau=0$ initial surface of the dS time slices. This operation is performed with the evolution operator generated by the $t$-Hamiltonian, the generator of translations in the foliation by $t=$ constant hyper-surfaces,   ${\widetilde H} \sim i\pt_t$.  Once the state is defined at $\tau=0$,
we may `push-forward' this state to any $\tau={\rm constant}$ surface in the dS patch, acting with the $\tau$-Hamiltonian $H \sim i\pt_\tau$. 

\bigskip
\centerline{\epsfxsize=0.4\hsize\epsfbox{combi.eps}}
\noindent{\ninepoint\sl \baselineskip=2pt {\bf Figure 2:} {\ninerm
Domains of bulk AdS covered by the bulk Hamiltonian developments in dS time slicing (left) versus E time slicing (right), for the same interval of boundary data. Notice that both domains share the initial-value surface $t=\tau=0$. The wavy lines represent a perturbative particle-like state which can be propagated smoothly to all values of the $t$ variable. }}
\bigskip

\subsec{Extracting UV Data}

\noindent

The `pull-back/push-forward'  method described here (to follow the terminology of \refs\susspull), provides a simple operational definition of `horizon complementarity' in a very concrete  example. As it stands, the construction applies to  perturbative states around the vacuum AdS manifold. 

In seeking generalizations, it is natural to look at the asymptotic (UV) data, whose non-perturbative CFT interpretation is most straightforward. 
In this vein, we look for the effect on the AdS boundary of the alternative Hamiltonian foliations in the bulk and pick a natural map between $\tau={\rm constant}$ and $t={\rm constant}$ surfaces to represent the complementarity.  

\bigskip
\centerline{\epsfxsize=0.3\hsize\epsfbox{pullpush.eps}}
\noindent{\ninepoint\sl \baselineskip=2pt {\bf Figure 3:} {\ninerm
Any state specified by bulk data on fixed $t$ surfaces may be mapped unitarily into a state at fixed $\tau$ by pulling it back to $t=\tau=0$
as an intermediate step. The matching of time slices at the AdS boundary defines the time-diffeomorphism $t_\tau$. 
}}
\bigskip

  Directly matching the fixed-$t$ and fixed-$\tau$ surfaces at the AdS boundary (cf. Figure 3) provides such a natural map, determining a particular time-diffeomorphism which we shall denote $t_\tau$. 
We can find its explicit form using the common $SO(d)$ symmetry of \eframe\ and \dsframe\ to set $d\Omega_{d-1} =0$ in  both metrics.   Introducing coordinates $s= 
\tan^{-1} (r) - 
\pi/2$ and $u_\pm = 
\shalf (t \pm s )$ we obtain the metric of the AdS$_{2}$ section of \eframe: 
\eqn\muno{
ds_{1+1}^2 = -4{ du_+\, du_- \over \sin^2 (u_+ - u_-)}\;.
}
If instead we define $v_\pm = \shalf(\tau \pm \eta)$ with 
$$
\eta = \half \log\left({\cosh(\rho) -1 
\over \cosh(\rho) +1} \right) \;,
$$
we find a metric
\eqn\mudos{
ds_{1+1}^2 = -4{ dv_+ \,dv_- \over \sinh^2 (v_+ - v_-)}
}
for the $d\Omega_{d-1} =0$ section of \dsframe. By direct inspection, we can check that \muno\ and \mudos\ are related by the
transformation
\eqn\diftt{
\tan (u_\pm) = \tanh (v_\pm)
}
on their  domain of overlap. This includes 
 the AdS boundary, defined by $u_\pm = t/2$ and $v_\pm = \tau/2$. On this boundary, the  diffeomorphism \diftt\ reduces to the sought-for time-map:
\eqn\edst{
\tan(t_\tau/2) = \tanh(\tau/2)
\;,
} or, equivalently $
\cos (t_\tau) = 1/\cosh(\tau)$. The result is of course consistent with the conformal map between boundary dS$_d$ and E$_d$ metrics \conf.

Associating unitary evolution operators to the two time foliations we may write
\eqn\pp{
|\tPsi\ket_{t_\tau} =   {\widetilde U}_{t_\tau} \;U_\tau^{-1} \,|\Psi\ket_\tau\;
}
for the unitary map between the two states at fixed $t$ or fixed $\tau$ respectively. Our particular matching of time foliations, given by the diffeomorphism $t_\tau$, allows us to interpret \pp\ as the unitary implementation of the conformal map $\Omega$ between E$_d$ and dS$_d$, i.e. 
\eqn\uconf{
U_{\Omega} =    {\widetilde U}_{t_\tau} \,U_\tau^{-1} \;, \qquad \cos (t_\tau) = {1\over \cosh (\tau)}
\;.
}
Notice that $U_\Omega$ acts on the Hilbert space at a given value of the either time parameter\foot{ Alternatively, we may use the Heisenberg language, where the state is fixed conventionally as the $t=\tau=0$ state, and Hermitian operators representing observables are evolved unitarily in time. The two frames translate then into two operator algebras obtained from the action of the respective evolution operators on a given $t=\tau=0$ observable $A_0$. We have $
A(\tau) = U_\tau^{-1} \,A_0 \,U_\tau$, and ${\widetilde A}(t) = {\widetilde U}_t^{\;-1}\,A_0\,{\widetilde U}_t$. 
Again, the two operator algebras are unitarily related by the operator $U_\Omega ={\widetilde U}_{t_\tau} \, U_\tau^{\,-1}$.}, sending  a dS-frame state $|\Psi\ket_\tau$ at  dS time $\tau$ into an E-frame state $|\tPsi\ket_t$ with $t= t_\tau$. The singularity of this map at $t=\pm \pi/2$ does not translate into a physical singularity for perturbative states around the AdS vacuum. Those states  are perfectly smooth in the E-frame and may be continued for all values of $t\in {\bf R}$. The crucial issue of whether this smoothness is expected for more general states will be addressed in the next section.

The relation \pp\  was motivated by the geometry of the Hamiltonian flows in the AdS geometry,  and the evolution operators could be   constructed in  the low energy theory of the bulk,     describing perturbative states around the AdS vacuum manifold. However, the resulting operators are parametrized by time variables that make sense in the exact CFT, so it is natural to promote \pp\ as a non-perturbative definition of the `complementarity map'. 

\subsec{(In)Completeness}

\noindent

The same method can be implemented for the case of the Poincar\'e patch, 
\eqn\ppa{
ds^2_{\rm Poincare} = -y^2 \,dt'^{\,2} + {dy^2 \over y^2 } + y^2\,d\ell^2_{{\bf R}^{d-1}}\;,
}
defined for $t' \in {\bf R}$ and $y>0$.  The physics on this patch is codified by the Minkowski version of the CFT, i.e. picking a conformal boundary with metric
${\bf R}\times {\bf R}^{d-1}$, and the unitary complementarity map between \ppa\ and \eframe\ can be constructed as in \uconf, using the common $t'=t=0$ initial value surface.
 
An interesting example where this method {\it does not} work in a naive fashion is provided by the hyperbolic foliation of AdS:
\eqn\hy{
ds_{\rm hyp}^2 = -({\bar r}^{\,2} -1) \,d{\bar t}^{\,2} + {d{\bar r}^{\,2} \over {\bar r}^{\,2} -1} + {\bar r}^{\,2} \,d\ell^2_{{\bf H}^{d-1}}\;,
}
where the radial coordinate is defined in the domain ${\bar r} >1$ and the boundary metric is taken to be ${\bf R} \times {\bf H}^{d-1}$, the second factor being a $(d-1)$-dimensional hyperboloid. This time, the ${\bar t}=0$ surface does not cover the whole $t=0$ surface of the global manifold.  The null surface ${\bar r}=1$ is a horizon of a particular black hole solution with hyperbolic horizon geometry and Hawking temperature $T=1/2\pi$, which suggests that a situation similar to that of the eternal AdS black hole is at play
\refs\eternalmalda. Indeed, one can cover the complete initial value surface with the ${\bar t}=0$ section of  two hyperbolic patches of the form \hy, each one of them dual to the CFT living on  ${\bf R} \times {\bf H}^{d-1}$. 

The global AdS background is dual to the  CFT on the ${\bf S}^{d-1}$ vacuum, which can be regarded as an entangled state of the Hilbert spaces supported on each hemisphere of ${\bf S}^{d-1}$. Since a $(d-1)$-dimensional hemisphere is conformal to ${\bf H}^{d-1}$, let ${\cal U} $ denote the unitary operator implementing the map on the CFT Hilbert space. The global vacuum can be written then as an entangled state with data on  two copies of the hyperbolic CFTs (cf. \refs\mvr):
\eqn\vacent{
|{\rm VAC}_{{\bf S}^{d-1}} \ket = \sum_{E_{\rm hyp}} e^{-\pi E_{\rm hyp} } {\cal U}_L |  E_{\rm hyp} \ket_L \otimes {\cal U}_R | E_{\rm hyp} \ket_R\;,
}
where $E_{\rm hyp}$ is an energy eigenvalue of the CFT quantized on the ${\bf H}^{d-1}$ spatial manifold\foot{This correspondence contains subtleties for CFTs with scalars, which have negative mass-squared due to the conformal coupling to the negative scalar curvature of ${\bf H}^{d-1}$. When the theory is defined on a compact quotient of  ${\bf H}^{d-1}$ there are genuine tachyonic zero modes and non-perturbative instabilities (cf. \refs\rmagan)} (the sum in \vacent\ is symbolic, since the spectrum of hyperbolic energies is continuous). 
Operators on a single copy see the AdS vacuum as a mixed thermal state with the temperature
$T=1/2\pi$ of the hyperbolic AdS black hole. \foot{Notice that ${\bf R}\times {\bf H}^{d-1}$ is conformal to the static patch of dS$_d$. Hence we are consistent with the global E$_d$/dS$_d$ map, since two static dS patches are needed to cover the sphere ${\bf S}^{d-1}$ at $\tau=0$ in the global dS space. }

 \bigskip
\centerline{\epsfxsize=0.5\hsize\epsfbox{combisone.eps}}
\noindent{\ninepoint\sl \baselineskip=2pt {\bf Figure 4:} {\ninerm
Causal diagram of the Poincar\'e patch (left figure) and the hyperbolic patches (right figure) in AdS. Unlike the previous global representations of AdS$_{d+1}$, the line denoted $L$ is a true boundary component, rather than the origin of polar coordinates and points represent surfaces homeomorphic to ${\bf R}^{d-1}$ rather than spheres. For $d>1$ the two boundary components, denoted $L$ and $R$ in the figure,  define  subsets  of the complete conformal boundary  ${\bf R} \times {\bf S}^{d-1}$. 
For $d=1$ this picture gives a complete representation of the causal structure, where the two boundary  components $L$ and $R$ are truly disconnected. 
}}
\bigskip

An important comment regarding \vacent\ is that, while the ${\bf S}^{d-1}$ vacuum state $|{\rm VAC}_{{\bf S}^{d-1}} \ket$ of the CFT  should map smoothly to the global AdS geometry, the same cannot be said of each individual eigenstate of the hyperbolic Hamiltonian $|E_{\rm hyp} \ket$. As emphasized in \refs\mvr, such states are expected to harbor bulk singularities (akin to `firewalls' \refs\fire) on the  horizon of the hyperbolic patch. In the next section we shall add a simple classical argument in favor of this interpretation.

Complementarity maps from a left-right symmetric slicing in hyperbolic time ${\bar t}$  to some global E-frame slice, $t$,  can be specified by operators of the form 
\eqn\aco{
U_{\rm C}  ={\widetilde {U}}_{t}  \left( {\cal U}_L^{-1} U_{
\bar t}^{-1}  \otimes  {\cal U}_R^{-1} U_{\bar t}^{-1}  \right)
\;.
}
In this expression, the first factor pulls the fixed-${\bar t}$ state in the product hyperbolic CFT  back into the ${\bar t}=0$  slices, undoes the conformal map back to each left-right hemispheres and finally it pushes  the full ${\bf S}^{d-1}$ state forward in E-frame time $t$.  Notice, however, that $U_{\rm C}$ defined in \aco\  makes use of the two copies of the CFT on disjoint hyperboloids, and the resulting operator does not have a straightforward interpretation as a unitary representation  of a conformal map in the full CFT defined on ${\bf R}\times {\bf S}^{d-1} $.

\subsec{$d=1$}

\noindent

We note that the  $d=1$ case has interesting peculiarities.    The union of the back-to-back hyperbolic patches of AdS$_{1+1}$  coincides
with the bubble patch. Their boundary is dS$_1$, consisting of two disconnected lines, each one representing one static dS patch (cf. Figure 4).  The map between Poincar\'e and global frames also simplifies. We compute here for future use  the associated boundary  time diffeomorphism. 
Let the Poincar\'e patch of AdS$_{1+1}$ be represented by the metric
\eqn\pin{
ds^2_{\rm Poincare} = -y^2 \,dt'^{\,2} + {dy^2 \over y^2}\;,
}
with $y\geq 0$ and $t' \in {\bf R}$, covering a proper subset of the global AdS$_{1+1}$ whose metric  we write as 
\eqn\glod{
ds^2_{\rm global} = -(1+x^2) \,dt^2 + {dx^2 \over 1+x^2} \;,
}
with $x \in {\bf R}$ and $t\in {\bf R}$, the right and left boundaries corresponding to the limits $x\rightarrow \pm \infty$ respectively.  As indicated  in Figure 4, a natural time-diffeomorphism $t_{t'}$ is induced on the boundary metrics by the matching of time slices at the R boundary $x=y=+\infty$. To find this boundary diffeomorphism we begin by transforming \pin\ by the change of variables 
$\zeta_\pm = t' \pm 1/y$, leading to
\eqn\pg{
ds^2 = -4\,{d\zeta_+ \,d\zeta_- \over (\zeta_+ - \zeta_-)^2}\;,
}
which in turn may be transformed into the global version \muno\ under the further redefinition $\zeta_\pm = \tan(u_\pm)$. Evaluating the chain of coordinate changes at the R boundary, we find 
\eqn\fc{
t' = \tan (t_{t'}/2)\;
}
for the required time-diffeomorphism. 

It is tempting to promote the picture of complementarity maps outlined in this section to conformal maps in CFT$_1$, i.e. a model of conformal quantum mechanics which would encode the physics of AdS$_{1+1}$ spaces. On the other hand, the $d=1$ version of the AdS/CFT correspondence is rich with subtleties (cf. for instance \refs{\frag, \sen}) which makes it a rather special case. Despite these caveats, we will find in the coming sections that many aspects of the complementarity maps discussed here do find analogs in the simplest models of conformal quantum mechanics.

\newsec{Singular Maps Versus Singular States}

\noindent 

We have argued that a version of horizon complementarity for perturbative bulk states around the global AdS vacuum can be analyzed in terms of conformal maps between the E$_d$ and dS$_d$ versions of the dual CFT. This conformal rescaling, which we refer to as the EdS map, sends the whole Hamiltonian development of the dS  manifold into a compact domain of Einstein-frame time. We refer to this situation as the `eternal/apocalyptic duality'. Accordingly, we speak of the `eternal Hamiltonian', dual to the dS time variable, $\tau$, and the `apocalyptic Hamiltonian', dual to the E-frame time variable, $t$. The conformal transformation $U_\Omega$ is singular at the endpoints of apocalyptic time $t=t_\star=\pm \pi/2$, but the physics of perturbative states around AdS is smooth, as the E-frame Hamiltonian acts smoothly on those states for $|t|> \pi/2$. 

It is possible to envisage states without such a  smooth continuation, for which the apocalyptic time development  is truly singular in a physical sense. Let us consider a classical state with the properties of a codimension-one brane,  supported on a fixed  
$\rho$ trajectory in \dsframe. Such a state is stationary with respect to the
$\tau$-Hamiltonian, but it is  accelerating, asymptotic to a null surface,   in the E-frame of the global AdS geometry. Therefore it requires an infinite supply of $t$-energy, and its $t$-time evolution is not expected to be smooth  for $\Delta t > \pi$. An example of this behavior
is given by a $O(d,1)$-invariant configuration similar to a  Coleman-de Luccia  (CdL) bubble, which expands exponentially in an ambient AdS space and produces a crunch as in Figure 5. \foot{These solutions can be constructed as Lorentzian continuations, in the sense of Hartle and Hawking \refs\hartleh, of $O(d+1)$-invariant Euclidean solutions with the interpretation of renormalization-group flows of the Euclidean CFT on  ${\bf S}^d$ (cf. \refs\usd).}

Brane-like states producing crunch singularities are a rather more interesting arena where ideas of complementarity can be    probed. Since the whole space-time crunches, they behave in some sense as infinite-entropy limits of black holes --even the boundary of AdS `crunches' in finite global time.  Local observables associated to constant-$\rho$ trajectories are analogous to `exterior' black hole observables, whereas local observables associated to constant-$r$ trajectories are analogous to `infalling' observables. Unlike the black hole case, we can identify infalling `observers' even on the AdS boundary, so that the complementarity map must be visible in the deep UV data of the CFT. 

 \bigskip
\centerline{\epsfxsize=0.2\hsize\epsfbox{bubblep.eps}}
\noindent{\ninepoint\sl \baselineskip=2pt {\bf Figure 5:} {\ninerm
A  $O(d,1)$ invariant bubble of finite dS energy, producing a crunch at $t=\pm \pi/2$. Surfaces of fixed $t$ in the exterior AdS geometry are indicated in the picture. If the shell is very thin, the interior geometry is also well approximated by AdS except near the bang/crunch singularities.
}}
\bigskip

 It is precisely the conformal transformation between `eternal' and `apocalyptic' Hamiltonian flows  what provides this    `UV remnant' of the complementarity map, visible in the microscopic formulation of the CFT.  
In what follows, we study the transformation between eternal and apocalyptic Hamiltonians from various points of view, starting with a Landau--Ginzburg description of the codimension-one brane states.

\subsec{Effective Landau--Ginzburg Models}
\noindent

          An approximate  description of  $O(d,1)$-invariant brane states can be achieved by defining a radial collective coordinate $\phi$ which can be regarded as a field degree of freedom in the CFT. Assuming that this world-volume field is weakly coupled, it can be assigned a   canonical mass  dimension. A brane situated at $\rho=\rho_M$ can be expressed by arranging the effective dynamics such that the collective field
$\phi$ obtains an expectation value
\eqn\conds{
\bra \phi \ket_M \sim M^{d-2 \over 2}\;,
}
where $M$ is the mass scale associated to the fixed radial position $\rho=\rho_M$. According to the IR/UV relation of AdS/CFT 
we have  (cf. \refs\usd) 
\eqn\uvird{
\rho_M 
\sim 
\log \,\bra \phi\ket_M \sim \log (M)\;,
}
a relation which is valid provided $d >2$ and  $M\gg 1$ in units of the dS curvature radius, two conditions that we assume to be valid throughout this section.

The simplest effective dynamics supporting such a classical condensate on dS is given by the effective (long wavelength) Landau--Ginzburg (LG) action 
\eqn\effl{ 
S[\phi]_{\rm eff} = -\int_{{\rm dS}_d}\left[ \shalf |\pt \phi |^2 + V_{\rm eff} (\phi) \right] \;
,}
where the effective potential can be written as 
\eqn\modlg{
V_{\rm eff} [\phi] = 
 \shalf \xi_d\, {\cal R}_{{\rm dS}_d} \,\phi^2 + \lambda\,\phi^{2d \over d-2} + \varepsilon M^{d-\Delta} \phi^{2\Delta \over d-2} \;.
}
The  first term and the marginality of the operator appearing in the second term are dictated by conformal invariance\foot{These terms are induced at large $\rho_M$ from the Dirac--Born--Infeld action of the brane (cf. \refs{\seiwit, \usu})}, including the conformal curvature coupling with 
$$
\xi_d = {d-2 \over 4(d-1)}\;.
$$
The non-linear terms in \modlg\ correspond to a marginal operator of mass dimension $d$ and a relevant operator of dimension $\Delta < d$, whose coupling introduces the conformal symmetry-breaking  scale $M$. The factor $\varepsilon = \pm 1$ controls the sign of the relevant operator, and we must require $\lambda >0$ for global stability.  In general, there may be many relevant operators and a host of irrelevant operators correcting \effl, but the simplified form of \modlg\ will suffice for our  qualitative discussion.

Taking $\lambda = \CO(1)$ and $M\gg 1$ we can find condensates of the form \conds\ provided $\varepsilon =-1$. In fact, we get both a stable condensate and an unstable one, as shown in Figure 6. The unstable condensate was interpreted in \refs{\usu, \usd} as  a sphaleron configuration which all the properties of a CdL bounce in the bulk. Interestingly, this configuration is present even for the globally unstable model with no relevant operator,   $M=0$ and $\lambda<0$. Such models  were studied extensively in \refs{\her, \craps,  \mald, \har, \usu, \usd} as holographic duals of crunch singularities. It was recognized in \refs{\mald, \harsus,  \usd} that
the stable condensates in globally well-defined models are perfectly suited to  the AdS/CFT embedding of space-times with crunch singularities.

The classical LG description of condensate states on dS should be accurate when the scale of the condensate is much larger than the dS temperature, i.e. $M\gg 1$ in our notation. In the opposite limit, $M\ll1$, the effective LG theory should receive large quantum corrections. On the other hand, this is the limit where classical gravity descriptions in the bulk admit a linearized approximation (cf. the appendix of \refs\mald), the result being $O(d,1)$-invariant geometries with very small bubbles and the same crunching behavior as in Figure 4

 \bigskip
\centerline{\epsfxsize=0.4\hsize\epsfbox{dspote.eps}}
\noindent{\ninepoint\sl \baselineskip=8pt {\bf Figure  6:} {\ninerm
Schematic representation of the de Sitter effective LG potential with $M\gg 1$ and $\varepsilon =-1$. The unstable condensate at the local maximum is dual to a CdL bubble in the bulk. }}
\bigskip

The conformal complementarity (EdS) map  \uconf\ becomes particularly simple in the classical approximation to the LG dynamics \effl. Given the conformal map between the two frames
\conf, an extension to the full effective LG field dynamics is achieved by postulating the conformal transformation of  the basic field variable, as dictated by its scaling dimension: 
\eqn\mad{
\tphi(t)= \Omega(t)^{d-2 \over 2} \,\phi(\tau_t)\;.
}
This transformation sends the dS-invariant condensate $\bra \phi \ket_M \sim M^{d-2 \over 2}$ into the $t$-dependent E-frame configuration 
\eqn\madd{
\bra \tphi (t)\ket_M  \propto \left({M \over \cos (t)}\right)^{d-2 \over 2}\;,
}
which is a solution of the E-frame system
\eqn\efsys{
{\widetilde S}_{{\rm E}_d} = \int_{{\rm E}_d} \left( \shalf |\pt \tphi |^2 + \shalf \xi_d {\cal R}_{{\rm E}_d} \,\tphi^{\,2}+ {\widetilde V}_{\rm eff} (\tphi\,) + \dots\right) \;,
}
with an effective potential 
 \eqn\epof{
 {\widetilde V}_{\rm E} [\tphi\,] = \shalf \xi_d\,{\cal R}_{\rm E} \,\tphi^2 + \lambda \,\tphi^{\,{2d \over d-2}} -\left({M \over \cos(t)}\right)^{d-\Delta} \,\tphi^{\,{2\Delta \over d-2}}\;
 }
 featuring an explicit time-dependent coupling of the relevant operator.  This coupling causes the 
  total energy, as well as the kinetic and potential energies of the state $\tphi(t)$ to blow up at the `bang-crunch' times $t_\star = \pm \pi/2$, showing that the singularities at the `apocalyptic' times are physical in terms of the E-frame variables. The E-frame Hamiltonian is itself singular at $t=t_\star$, so that the $t$ time  evolution cannot possibly continue smoothly beyond the apocalyptic times. 

We conclude that a  particular class of $O(d,1)$-invariant  states in dS-frame variables  are seen as a singular (crunching) states in the E-frame, as a result of a singular driving term in the E-frame Hamiltonian. 

By inverting \mad\ we can study how an E-frame $t$-stationary  condensate looks when analyzed in dS-frame variables. Such states have $U(1)\times O(d)$ symmetry and have the form $\bra \tphi\,\ket_{\widetilde M}  \propto {\widetilde M}^{\,{d-2 \over 2}}$  (notice that we now take $\pt_t {\widetilde M} =0$). This  $t$-static configuration is a solution of the static E-frame potential
$$
{\widetilde V}_{\rm E} [\tphi] = \shalf \xi_d\,{\cal R}_{\rm E} \,\tphi^2 + \lambda \,\tphi^{\,{2d \over d-2}} - {\widetilde M}^{\,d-
\Delta} \,\tphi^{\,{2\Delta \over d-2}}\;.
$$
  The corresponding dS-frame field is
\eqn\maddd{
\phi_{\widetilde M} (\tau) \propto \left( {{\widetilde M} \over \cosh (\tau)}\right)^{d-2\over 2}\;, 
}
which vanishes exponentially in global dS time for $d>2$.  After appropriately normalizing the $\CO(1)$ proportionality constant in \maddd, this solution is driven by the dS-frame LG `potential' 
\eqn\dslg{
V_{\rm dS} [\phi] = \shalf \xi_d \,\CR_{{\rm dS}_d} \,\phi^{\,2} + \lambda \,\phi^{\,{2d\over d-2}} - \left({{\widetilde M} \over \cosh(\tau)}\right)^{d-\Delta} \,\phi^{\,{2\Delta \over d-2}}\;,
}
which now features a negative-definite, $\tau$-dependent relevant operator turning-off as $|\tau| \rightarrow \infty$. The value of
the LG potential evaluated on the solution \maddd\  also redshifts to zero as $|\tau|\rightarrow \infty$. 
We thus conclude
that the $U(1)\times O(d)$-invariant  condensates on the E-frame  `dilute away' when analyzed in dS-frame variables. 

Broadly speaking, we can identify two qualitatively different types of states.
One natural class is given by those states which are completely smooth in the E-frame and can be continued through all $t\in {\bf R} $ with a time-independent non-singular E-frame Hamiltonian. We refer to these as {\it smooth} states. When analyzed in the eternal frame, their distinctive feature is the `diluting' nature as $|\tau| \rightarrow \infty$. 

A second class of states is given by those which are asymptotically $\tau$-stationary in the eternal frame, but distinct from the trivial CFT vacuum on dS. The natural way of engineering such states is to deform the CFT by a relevant operator and consider stationary states looking like condensates induced by the new relevant interactions.
These states, while completely regular in the eternal dS frame, are singular in the E-frame and therefore called {\it crunch} states.  

It should be clear that the smooth and crunch states do not share the same phase (or Hilbert) space. They actually occur in different systems, in the sense that they need different Hamiltonians to be supported as stationary states. If we fix, say the dS frame, crunch states need a non-trivial dS-invariant relevant deformation to be turned on, while smooth states already exist in dS systems whose Hamiltonian has no such deformation turned on. 

We have chosen to discuss the conformal map which rises naturally from the diffeomorphisms discussed in section 2.
It maps a non compact region into a compact one independent of the dynamics brought about by the specific
Hamiltonian involved. This more universal approach required us to disentangle the singularity inherent in such a transformation from a possible dynamical one. One could have chosen a conformal transformation akin to a unitary
gauge in gauge theories.  It would be {\it ab initio} useful when there is a physical singularity to be exposed in one frame,
like the unitary gauge is useful in the BEH phase. The transformation will be defined on the fields
(cf.  equation \mad) in such a way that  $ \Omega$  is multiplied by  the product of the expectation value of
the scalar field $\phi$ in the dS frame and the Hubble scale. This product  vanishes in the cases when there is no condensate and thus renders the transformation to be ill defined in those cases.

\subsec{Classical Firewalls}

\noindent

It is interesting to inquire to what extent this  description generalizes to other versions of the conformal frame duality studied here, such as the examples of AdS foliations related to CFTs on flat or hyperbolic space-times.  

Let ${\bf K}_k$ represent the standard constant-curvature manifold in $d-1$ dimensions, with $k=0, \pm 1$ controlling the sign of the Ricci curvature, i.e. ${\bf K}_0 = {\bf R}^{d-1}$, ${\bf K}_1 = {\bf S}^{d-1}$ and ${\bf K}_{-1} = {\bf H}^{d-1}$.   We can describe the global,  Poincar\'e and hyperbolic patches of AdS at once with the family of metrics:  
 \eqn\manyfol{
ds^2_k = -(r_k^2 + k)\,dt_k^2 + {dr_k^2 \over r_k^2 + k} + r_k^2 \,d\ell^2_{{\bf K}_k}\;.
}
The  $k=1$ case with $r_1 \geq 0$ 
is the standard metric of the global AdS manifold \eframe. The case  $k=0$ with $r_0 \geq 0$  gives the Poincar\'e patch \ppa\ of AdS, and finally $k=-1$  with $|r_{-1}| \geq 1$ returns 
 the two mirror hyperbolic patches given by  \hy. The time variables $t_k$ in 
\manyfol\ define natural Hamiltonian flows for CFTs on ${\bf R} \times {\bf K}_k$. In the notation of the previous section, we
have $t=t_1$, $t' = t_0$ and ${\bar t}= t_{-1}$.

   It is interesting to inquire about the fate of condensate states with the symmetries of ${\bf R} \times {\bf K}_k$, corresponding to brane-like states defined by $r_k = {
   \rm constant} $ in \manyfol. In particular, one can consider condensates on ${\bf R}\times {\bf R}^{d-1}$ with Poincar\'e invariance $ISO(d-1, 1)$ and condensates on ${\bf R} \times {\bf H}^{d-1}$ with symmetry $U(1) \times O(d-2, 1)$. \foot{Interestingly, these condensates can be defined without the need of a relevant operator. The $k=0$ flat-space case is similar to a Higgssed state on a Coulomb branch and the $k=-1$ hyperbolic case only requires the existence a positive marginal deformation to stabilize the negative-definite conformal mass term.}

The crucial property making the $k=0$ and $k=-1$ cases special is the non-compact nature of the spatial section ${\bf K}_k$. Unlike the EdS map studied so far, this implies that the conformal transformation to the E-frame:
\eqn\cofw{
ds^2_{k=0, -1} = \Omega(x)^2 \,ds^2_{{\rm E}_d}\;,
}
has singularities even on the $t=0$ spatial section, at those points on ${\bf S}^{d-1}$ where the infinite boundary of ${\bf K}_k$ is mapped.  In particular, a maximally symmetric condensate on ${\bf R} \times {\bf K}_k$ of the form
\eqn\codm{
\bra \phi \ket_k \propto M^{d-2 \over 2}
}
is mapped to an E-frame field 
\eqn\esin{
\bra \tphi(x) \ket_k \propto \left(\Omega(x) M \right)^{d-2 \over 2} \;,
} 
with nontrivial space-time profile, and sharing  the  singularities of the Weyl function $\Omega(x)$. 
The  configuration \esin\  solves the E-frame effective equations of motion with a relevant perturbation
\eqn\srcc{
{\widetilde V}_\Delta [\tphi] = - \left(M \,\Omega(x)\right)^{d-\Delta} \,\left(\tphi(x) \right)^{\, {2\Delta \over d-2}}\;.}
 The physical interpretation in the E-frame is that of an inhomogeneous injection of energy with sharp divergences at the singularities of the conformal map. This happens for $k=0$ at a single point on ${\bf S}^{d-1}$, whereas the infinite injection of energy occurs in the $k=-1$ case along the complete equatorial ${\bf S}^{d-2}$ which separates ${\bf S}^{d-1}$ into two hemispheres. It follows that  singularities of `firewall' type are expected in the global bulk description of such states, in agreement with the philosophy  expressed in \refs\mvrr. The price we pay for the ability to use a classical set up is the need to realize the state in a CFT perturbed by a large relevant operator, but the take-away message ends up being the same.

The behavior of homogeneous condensate states described in this section should admit a natural extension for small perturbations around these states, such as finite-particle excitations. On the other hand, it would be
interesting to generalize the present purely classical description to the full quantum theory. The presence of strongly time-dependent couplings makes the problem challenging. Fortunately,  a number of structural properties of the complementarity maps can be studied in a simplified quantum mechanical model, where time-dependent couplings can be studied at considerable ease.

\newsec{Conformal Quantum Mechanics}

\noindent

In order exhibit these ideas in an explicit quantum framework we can study the quantum mechanical version of
the Landau--Ginzburg models associated to conformal complementarity maps.  A natural construction arises as the $d\rightarrow 1$ limit of the above, in which we replace the classical $d$-dimensional  conformal dynamics of the LG collective degree of freedom  by its $d=1$ analog. It turns out that this simple procedure is somewhat non-trivial, since the different frames  will be found to retain some characteristic features in the $d=1$ system.  

The basic building block is given by the de Alfaro--Fubini--Furlan (AFF) Conformal Quantum Mechanics (CQM) with
Hamiltonian \refs\aff\
\eqn\daff{
 H(\pi, \phi)_{\rm AFF} = \half\left(\pi^2 + {\lambda \over \phi^2}\right)\;,
 }
 for one LG-type degree of freedom $\phi$ with canonical momentum $\pi$.  
  The conformal group acts on the Hilbert space of this theory as an $SL($2,{\bf R}$)$  algebra generated by the Hamiltonian \daff, the dilatation operator $D= \shalf \{\phi, \pi\}$ and the special-conformal generator $C= \shalf\phi^2$, with commutation relations
 $$
 [D, H] = 2iH\;, \qquad [D,C]=-2iC\;,\qquad [H,C] = -iD\;,
 $$ 
 which follow from the basic canonical Heisenberg algebra $[\phi, \pi] = i$. 
 
 The AFF Hamiltonian is classically bounded-below for  repulsive potentials with $\lambda>0$. Even when the potential 
becomes attractive, it remains well defined at the quantum level as long as $\lambda > -1/4$. The spectrum is still well defined for $\lambda > -1/4$ when the system is quantized on $L^2 ({\bf R}^+)$, i.e. on wave-functions $\Psi[\phi]$ with inner product
 $$
 \bra \Psi | \Phi \ket = \int_0^\infty d\phi\,\Psi^* [\phi] \,\Phi[\phi]\;
 $$
 and vanishing boundary condition at the origin, $\lim_{\,\phi\to 0} \Psi[\phi] =0$. 
 More specifically, the Hamiltonian has a positive-definite continuous spectrum \foot{In the absence of a scale, no bound states form.} with delta-function normalization for  
 $-1/4 < \lambda $.
 
  A discrete spectrum can be obtained by placing the system on a   `harmonic trap', i.e. by adding a harmonic potential term with some frequency $\omega$, 
 \eqn\trap{
 H_\omega = H_{\rm AFF} + \omega^2\,C\;,
 }
 where $C = \shalf \phi^2$ is the generator of special conformal transformations. The main advantage of this IR regularization is the preservation of a  nice $SL(2,{\bf R})$ action  on the spectrum, since the Hamiltonian is still linear in the $SL(2,{\bf R})$ generators.  This leads in particular to an equally-spaced discrete spectrum for the trapped Hamiltonian $H_\omega$. 
 
 The trapped models  are analogous to the higher-dimensional conformal field theories defined on spheres, with a gapped spectrum, i.e. the model referred above as the E-frame CFT.  
 The analogy can be sharpened by doing  `dimensional reduction', namely taking the $d\rightarrow 1$ limit of \modlg. The conformal mass term does survive
this limit. The curvature's vanishing is compensated by  the behavior of the conformal coupling $\xi_d$, the product leading to a finite result. Explicitly, one finds for the LG model on ${\bf X}_k = {\bf R} \times {\bf K}_k$
\eqn\trapdr{
\omega_k^2 = \lim_{d\to 1}\, \xi_d \,{\CR}_{{\bf X}_k} = \lim_{d\to 1} \, {d-2 \over 4(d-1)} \,k\,(d-1)(d-2)= {k\over 4} \;
,}
and for the LG model on dS$_d$: 
\eqn\dslimdr{
\lim_{d\to 1}\, \xi_d \,{\CR}_{{\rm dS}_d} = \lim_{d\to 1} \,{d-2 \over 4(d-1)} \,d(d-1) = -{1\over 4}\;.
}
It is interesting that we get the same  {\it tachyonic} `anti-trapping' frequency for the  $d\rightarrow 1$ limits of the hyperbolic and dS theories. This result is natural given the interpretation of the LG models as world-volume descriptions of codimension-one branes on AdS, since we have seen in section 2 that hyperbolic and `bubble' patches of AdS are identical for $d=1$. 

The complete LG action can be derived following the logic  of \refs\kallosh. We can drop a particle probe of mass $m$ in AdS$_{1+1}$ and analyze its near-boundary, slow-motion dynamics in each of the relevant patches:
$$
ds^2_{(k)} = -(r_k^2 + k)\,dt_k^2 + {dr_k^2 \over r_k^2 + k}\;,
$$
in the notation of \manyfol. The particle action reads
$$
S_{(k)} = -m\int dt_k \sqrt{r_k^2 + k -{1 \over r_k^2 + k} \left({dr_k \over dt_k}\right)^2 }\;,
$$
and takes the form of a CQM system with parameters $\omega_k^2 = k/4$ and $\lambda = 2m^2$:
\eqn\cqmk{
S[\phi_k] = \half \int dt_k \left[\left({d\phi_k \over dt_k}\right)^2 - \omega_k^2 \,\phi_k^2 - {\lambda \over \phi_k^2} \right]
}
	in the limit $r_k \gg 1$ and $|dr_k /dt_k  | \ll 1$, where we have used the field redefinition \foot{The canonical mass dimension of the field variable is $-1/2$ in $d=1$, so that the UV/IR relation between $\phi$  and the AdS radius is inverted,  the asymptotic AdS region corresponding now  to the small-field regime.}
\eqn\fruno{
\phi(t_k) = \left({4m \over r_k (t_k)}\right)^{1/2}\;.
}

We may thus consider three different versions of the CQM model. The standard AFF model with $\omega^2 =0$ (no trapping) 
will be regarded as the analog of  the M-frame, whereas the model with positive trapping $\omega^2 = 1/4$ corresponds to the E-frame. Finally,  the model
with tachyonic anti-trapping $\omega^2 = -1/4$ will be interpreted as the dS-frame (or equivalently the hyperbolic frame).  

Although the probe-brane derivation is very transparent, its logical relation to a well-defined AdS$_2$/CFT$_1$ duality is still far from clear. The asymptotic boundary conditions in AdS$_2$ are very sensitive to back-reaction from any finite-energy perturbation \refs\frag\ and the most likely interpretation of the AdS$_2$/CFT$_1$ correspondence involves a large Hilbert space with exactly degenerate states on the CFT side \refs\sen, whose precise relation to AFF-like models is an open problem. We shall not deal with such subtleties in this paper, our aim being more modest. Namely we use the AFF model as a quantum arena to study the eternal/apocalyptic map, while at the same time offering a tentative bulk interpretation of the results. 

In this respect, it is interesting to notice that the three basic versions of CQM can be entirely determined from the boundary time diffeomorphisms
discussed in section 2. To see this, consider the generator of time translations in the AFF model (aka the M-frame Hamiltonian), $H_{\rm M} = i\pt_{\bar t}$. We can realise the $SL(2, {\bf R})$ group by adding the generators $D=-2i \,{\bar t}\, \pt_{\bar t}$ for dilatations and $C= i {\bar t}^{\;2} \,\pt_{\bar t}$ for
special conformal transformations. Defining then $H_{\rm E} = H_{\rm M} + \sfourth C = i\pt_t $ and $H_{\rm dS} = H_{\rm M} - \sfourth C = i\pt_\tau$, the associated  differential equations are solved by the same diffeomorphisms, \edst\ and \fc, defined in section 2,  under the simple rescaling ${\bar t} = 2t'$, which is itself a symmetry of the AFF model.\foot{We thank the referee for suggesting this argument.}

 \subsec{Deformations And Bound States}
 
 \noindent

More generally, we can deform the AFF model (either trapped or untrapped) by adding a relevant operator contributing to the potential energy as
\eqn\relef{
V_{\Delta} (\phi) = \varepsilon \,{M^{1-\Delta} \over \phi^{2\Delta}}\;,
}
with $\Delta < 1$ (the trapping harmonic term being the particular case $\Delta = -1$). 

We notice that positive  relevant deformations with $\varepsilon >0$ and $\Delta <0 $ behave qualitatively like the trapping term \trap, in the sense that they remove
all the large-$\phi$  `scattering states' near zero energy. Hence, we interpret the models with such a strongly relevant deformation as completely gapped. 
For $\Delta =-1$ we have the strict harmonic trapping, analogous to the E-frame CFT. For $\Delta <-1$ they present a steeper wall, mimicking  a confining potential with a gap proportional to $M$ as $M\gg 1$. Since the complete large-$\phi$ region is removed from the spectrum,  we suggest to interpret such `confining' models as analogous to a sharp wall where  AdS$_2$ is terminated, as in a `bubble of nothing' \refs{\wbn, \dsconf}. 

On the other hand, relevant deformations in the window $0 < \Delta < 1$ are very mild at large values of $\phi$, preserving the continuum of large-$\phi$ scattering states near zero energy. Therefore, we interpret these deformations as leaving behind a sort of `IR CFT fixed point', such as the effective field theory describing  the IR behavior of a system where spontaneous symmetry
breaking has occurred.  In particular,
for $\varepsilon <0$ and large $M$ we find localized classical ground states at $\bra \phi \ket \sim 1/\sqrt{M}$ which we may identify as `condensate' states (cf. figure 7). Such states are analogous to codimension-one brane states propagating in AdS$_2$. 

The $\Delta =1$ case is the marginal deformation. Interestingly, a negative $\varepsilon =-1$ deformation does not automatically imply a global instability of the model, reminiscent of the CdL solutions discussed in the classical models above. The reason is the improved quantum stability\foot{This quantum stability in the small $\phi$ region is further enhanced in the large $N$ limit of the generalization with $O(N)$ symmetry. A term encoding the $N$-dependence of the measure would stabilize the system at large $N$, even in   the absence of a stabilizing  marginal operator.}  which sets the critical value of the effective coupling at $\lambda_{\rm critical} = -1/4$, a phenomenon analogous to the limited tolerance of tachyons in AdS \refs\BF. 

The AFF model admits exact solutions for the condensate states for the particular case of a $\Delta = 1/2$ deformation, since  the resulting induced potential \relef\ is a standard Coulomb interaction. It follows that a spectrum of bound states can be constructed as the radial Hydrogen wave functions continued to real values of the angular momentum, i.e. as (hypergeometric) solutions of
\eqn\hydr{
\left(-\half {d^2  \over d\phi^2} + {\lambda \over 2\phi^2} - {\sqrt{M} \over \phi}\right) U_n (\phi) = E_n \, U_n (\phi)\,}
with discrete spectrum of energies 
\eqn\enerd{
E_n = -{2M \over (2n+1+\sqrt{1+4\lambda})^2}\;, \qquad n\in {\bf Z}_+\;.}
 
  \bigskip
\centerline{\epsfxsize=0.5\hsize\epsfbox{potentials.eps}}
\noindent{\ninepoint\sl \baselineskip=8pt {\bf Figure 7:} {\ninerm
The AFF  potential deformed by (a) a positive strongly relevant operator with $\Delta <-1$ (confinement), (b) a harmonic potential, $\Delta=-1$ (trapping),  and $({\rm c}) $
a negative, mildly relevant deformation, $0<\Delta<1$ (condensate). 
}}
\bigskip

The notion of condensate states is inherently semiclassical in the particular case of the dS-frame Hamiltonian, which we write here explicitly, 
$$
H_{\rm dS} = \half\left(\pi^2 + {\lambda \over \phi^2} \right) - {1\over 8} \,\phi^2 - {\sqrt{M} \over \phi}\;,
$$
perturbed by a negative $\Delta=-1$ operator.
The condensate state induced by the last term is 
 necessarily metastable (cf. Figure 8). If this metastable well is deepened  by going to large $M$, the decay probability to the quadratic runaway region is of order $\exp(-a \,M^{2/3})$ for some constant $a$. The bulk interpretation is that of a probe particle which can tunnel out of an accelerating fixed-$\rho$ trajectory, into the low radius region of AdS$_2$. Any such probe that tunnels back to the interior of AdS fails to reach the boundary with infinite E-frame energy, and thus the crunch is prevented. We will return to this intriguing question in section 6. 
 
  \bigskip
\centerline{\epsfxsize=0.5\hsize\epsfbox{dSmetas.eps}}
\noindent{\ninepoint\sl \baselineskip=8pt {\bf Figure 8:} {\ninerm
The dS-frame potential, with tachyonic anti-trapping (dashed line) and the same model with a relevant operator inducing a metastable condensate (full line).
}}
\bigskip

\newsec{The CQM Complementarity Map}

\noindent

We now study the `conformal complementarity' in the CQM model. For $d=1$, it reduces to the conformal transformation induced by the time-diffeomorphism 
$$
 dt = {d\tau \over \Omega(\tau)}
 $$
 acting as a map between `eternal' time evolution, $\tau \in {\bf R}$, and `apocalyptic' time evolution,  $t\in [-t_\star, t_\star ]$. 
At the classical level we seek the appropriate Weyl function $\Omega(t)$ which maps the E-frame version of CQM:
\eqn\apofa{
{\widetilde S} = \int dt  \left[\half \left({d\tphi \over dt}\right)^2 - {1\over 2} \tomega^{\,2} \tphi^{\,2} - {\lambda \over 2\tphi^{\,2}}\right]\;,
}
with $\tomega^{2} = 1/4$, into the two canonical models of `eternal' type:
\eqn\etfa{
S= \int d\tau \left[ \half \left({d\phi \over d\tau}\right)^2 - \half \omega^2 \,\phi^2 - {\lambda \over 2\phi^2}\right]\;,  }
where $\omega^2 =0$ for the M-frame CQM and $\omega^2 =-1/4 $ for the dS-frame CQM (we use the same time variable for both eternal models for simplicity of notation). 

The answer is obtained by direct substitution of the conformal rescaling $\phi(\tau) = \tphi(t) \, \sqrt{\Omega(t)}$ into the actions. We find the required behavior up to a boundary term: 
\eqn\btermsdu{
S = {\widetilde S} -\int d\tau {dK\over d\tau} = {\widetilde S} - \int dt {d{\widetilde K} \over dt}\;,
}
where\foot{In a slight abuse of notation, we shall often denote the functions $K(\tau) = {\widetilde K} (t)$ with the same letter and drop the twiddle  in ${\widetilde K} (t)$, just as we do with the Weyl factor $\Omega(\tau)={\widetilde \Omega}(t) \rightarrow \Omega(t)$.  } 
\eqn\bterms{
K(\tau) = -{1\over 4} \,\phi^2 \,\pt_\tau \log \Omega(\tau)\;, \qquad {\widetilde K}(t) = -{1\over 4} \, \tphi^{\,2} \,\pt_t \log \Omega(t)\;,
}
provided the Weyl function  satisfies 
\eqn\trasfr{
\tomega^{\,2} = \Omega^2 \, \omega^2 + \half \Omega \,\pt^2_\tau \Omega - {1\over 4} (\pt_\tau \Omega)^2\;,
}
a relation that we may interpret as an `anomalous' transformation law for the frequencies. It is useful to define 
\eqn\ano{
\CA\equiv \half \left( \Omega\, \omega^2 - \Omega^{-1} \,\tomega^{\,2} \right) 
}
for future use, as a measure of such anomalous scaling behavior. In this notation, \trasfr\ reads
\eqn\anos{
\CA = {1\over 8} \Omega^{-1} \left( \pt_\tau \Omega\right)^2 - {1\over 4} \pt_\tau^2 \Omega = {1\over 8} \Omega^{-1} \left(\pt_t \log \Omega \right)^2 - {1\over 4} \Omega^{-1} \pt_t^2 \log \Omega\;.
}

Plugging into   \trasfr\  the actual values of the frequencies, we find two solutions of the non-linear differential equation which, not surprisingly, exactly match the time diffeomorphisms \edst\ and \fc\ found in the context of purely geometrical considerations in AdS$_{1+1}$.

 We have the EM map between the E-frame and M-frame systems, i.e. between the trapped and ordinary AFF models:
\eqn\daffp{
{\rm EM}: \;\;\;\omega^2=0\;, \qquad  \tomega^{\,2} = {1\over 4}\;, \qquad \Omega_{\rm EM} = \half(1+\tau^2) = {1\over 2\cos^2 (t/2)}\;.
}
 The second solution is the standard EdS map, between the trapped and tachyonic versions of the AFF model: 
 \eqn\edsp{
{\rm EdS}:\;\;\;\omega^2 =-{1\over 4}\;, \qquad \tomega^{\,2} = {1\over 4}\;, \qquad \Omega_{\rm EdS} = \cosh(\tau) = {1\over \cos(t)}\;.
}
A useful parametrisation of the two Weyl functions at once is
\eqn\wft{
\Omega_\alpha (t) = {1\over \alpha} \left({1 \over \alpha \cos (t/\alpha)}\right)^\alpha
\;,}
where $\alpha =1$ for the EdS map and $\alpha=2$ for the EM map.

Notice that the singularities of $\Omega(t)$ occur at $t=\pm t_\star$ with $t_\star = \alpha \pi/2$, in agreement with the geometrical features of AdS$_{1+1}$ Penrose diagrams, showing that the Minkowski patch covers a larger portion of the AdS boundary as compared to the dS (hyperbolic) patch (cf. Figure 4).

A relevant operator deformation of the form 
\eqn\relo{
\int d\tau \,{M^{1-\Delta} \over  \phi^{2\Delta}} 
}
in the eternal frame transforms into an analogous term
\eqn\reloa{
\int dt \, {{\widetilde M}^{\;1-\Delta} \over \tphi^{\,2\Delta}}
}
in the apocalyptic frame, where the mass parameters are related by
\eqn\mast{
 {\widetilde M}  = \Omega\,M\;.
 }
 Notice that the map between \relo\ and \reloa\ works also for time-dependent mass parameters, and \mast\ implies that either
 $M$ or ${\widetilde M}$ must be time-dependent in one of two frames.

\subsec{Quantum  Map}

\noindent

The field redefinition between the eternal and apocalyptic frames is generalized to a full quantum map by a correspondence between wave functions
$$
\Psi[\phi, \tau] \longrightarrow \tPsi[\tphi, t]
$$
given by the explicit transformations
\eqn\canouno{
\tPsi[\,\tphi, t\,] = \Omega(t)^{1 \over 4} \,e^{i{\widetilde K}(t) } \,\Psi\left[ \tphi\sqrt{\Omega(t)}, \tau(t)\right]\;,
}
and its inverse 
\eqn\canodos{
\Psi[\,\phi, \tau\,] = \Omega(\tau)^{-{1 \over 4}} \,e^{-iK(\tau)} \, \tPsi\left[\phi/\sqrt{\Omega(\tau)}, t(\tau)\right]\;.
}
In these expressions, $\tau(t)$ and its inverse give the appropriate time diffeomorphism transforming the eternal and apocalyptic frames. The first factor in \canouno\ and \canodos\ is a Jacobian accounting for the correct normalization of both wave functions and the phase is the result of the boundary term in time \btermsdu. It can be checked explicitly that this  map sends solutions of the apocalyptic Schr\"odinger equation 
\eqn\schapo{
i\pt_t \tPsi[\tphi, t] = {\widetilde H}\left(\tphi, -i{\pt \over \pt\tphi}\right) \tPsi[\tphi, t]\;,
}
into solutions of  the eternal Schr\"odinger equation
\eqn\schete{
i\pt_\tau \Psi[\phi, \tau] = H\left(\phi, -i {\pt \over \pt \phi}\right)\,\Psi[\phi, \tau]\;,
}
and viceversa, where the two dual Hamiltonians are defined as 
\eqn\twohals{
H= \half \pi^2 + \half \omega^2 \,\phi^2  + V(\phi)\;, \qquad {\widetilde H} = \half {\widetilde\pi}^{\,2}  + \half \tomega^{\,2}\,\tphi^{\,2}  + {\widetilde V}(\tphi\,)\;.
}
with 
\eqn\defpis{
\pi= -i{\partial \over \pt \phi}\;, \qquad  {\widetilde \pi} = -i{\partial \over \pt \tphi}\;.
}

 The quantum map \canouno\ and \canodos\ is formally a canonical transformation 
 $$
 (\phi, \pi) \longrightarrow (\tphi, \tpi) = \left({1\over \sqrt{\Omega}} \;\tphi,  \sqrt{\Omega} \;\tpi\right)\;,
 $$
 a change of variables that holds in quantum averages up to some anomalous terms coming form the boundary terms. 
We have the basic rules
\eqn\genmapo{
e^{iK} \;\pi \;e^{-i K} = \Omega^{-{1 \over 2}} \left( {\widetilde \pi} + \shalf  \tphi \,\pt_t \log \Omega\right)\;, \qquad e^{-i \widetilde{K}} \;{\widetilde \pi} \;e^{i\widetilde{K}}= \Omega^{1 \over 2} \left(\pi -\shalf \phi\,\pt_\tau \log \Omega\right)
}
which allow us to formulate the general map of observables for general polynomial functions of the canonical operators. In average values defined by 
\eqn\averdef{\eqalign{
\left\langle F(\phi\;;\;{ \pi}\,)\right\rangle_{\Psi} &\equiv \int_0^\infty d\phi \,\Psi^*[\phi, \tau] \,F(\phi, -i\pt_\phi) \,\Psi[\phi, \tau] \;, \cr
\left\langle F(\tphi\;;\; {\widetilde \pi}\,)\right\rangle_{\tPsi} &\equiv \int_0^\infty d\tphi\,\tPsi^* [\tphi, t]\,F\left(\tphi, -i\pt_{\,\tphi}\right)\,\tPsi[\tphi, t]\;.
}}
we have 
\eqn\opemap{\eqalign{
\left\langle F(\tphi\;;\; {\widetilde \pi}\,)\right\rangle_{\tPsi} &= \left\langle F\left(\Omega^{-{1 \over 2}} \phi\;;\; \Omega^{1 \over 2} (\pi -\shalf  \phi \,\pt_\tau \log \Omega) \right)\right\rangle_{\Psi}
\cr
\left\langle F(\phi\;;\;{ \pi}\,)\right\rangle_{\Psi} &= \left\langle F\left(\Omega^{1 \over 2} \tphi\;; \;\Omega^{-{1 \over 2}} ({\widetilde \pi} +\shalf \tphi\, \pt_t \log \Omega) \right)\right\rangle_{\tPsi}
\;.}}
These two equations can be used to extract information about the behavior of any observable. 

A consequence of the quantum map defined above is the complementarity of time evolutions in the respective `eternal' and `apocalyptic' frames. In particular, we can explicitly check the non-commutativity of  the time evolution operators,
\eqn\teo{
{\widetilde U}_t= T_t \, \exp\left(-i\int_0^t dt'\,{\widetilde H(t')}\right)\;, \qquad U_\tau = T_\tau \,\exp\left(-i\int_0^\tau d\tau' \,H(\tau') \right)\;,
} 
 by directly showing that the respective Hamiltonians fail to commute, even at the initial surface where $\Omega=1$. We can compute $[H, {\widetilde H}]$ by expressing, say ${\widetilde H}$ in  eternal-frame variables as 
$$
{\widetilde H} = \half\,{\widetilde \pi}^{\,2} + \half \,\tomega^{\,2}\,\tphi^{\,2}  +{\widetilde V}(\,\tphi\,)= \half\Omega^{-1} \, \tomega^2 \,\phi^2 + \Omega \left(\half \pi^2+ V(\phi)\right)\;.
$$ 
Hence we have the identity
$$
\Omega^{-1} {\widetilde H}= H + \half \left( \Omega^{-2} \tomega^2 - \omega^2 \right) \phi^2\;,
$$ 
from which we find  the commutator of the Hamiltonian operators 
\eqn\comh{
\left[H, {\widetilde H}\right] = 2i\,\CA\,D = 2i\,\CA\, {\widetilde D}\;.
}
in terms of the function $\CA$ defined in \ano\ and \anos, and the dilation operator 
\eqn\dilo{
D= \shalf \{ \phi, \pi\} = \shalf \{\tphi, \tpi \} = {\widetilde D}\;.
}
Since $\CA \neq 0$ even for $\Omega=1$, the two Hamiltonians do not commute in general. Note that they {\it do} commute on those states which are annihilated by the generator of the scale transformations. 
 This interpretation makes contact with the fact that the vacuum AdS manifold realizes the boundary conformal group as an isometry group, showing that the bulk geometry is actually codifying the quantum state of the dual CFT. 

It is interesting to notice that the scale-invariant wave function, satisfying $D\Psi_0 =0$, is given by
$
\Psi_0 (\phi) =  \phi^{-1/2}
$
and is {\it not} normalizable, i.e. it does not sit in the Hilbert space of bound states. Still,  its norm  is less divergent than that of a plane wave. 

\subsec{States And Observables}

\noindent

The quantum map \canouno\ and its inverse \canodos\ are completely general, valid for any state with arbitrary wave-function.  Both versions of the complementarity map have a Weyl function which smoothly tends to the identity  near the origin of times, $\Omega =1$ for $t=\tau=0$, and blows up at the `ends of time',  with a pole-like behavior  $\Omega(t) \sim (t-t_\star)^{-\alpha}$ near $t=\pm t_\star = \pm \alpha\pi/2$.
 
In order to further fix the intuition about the meanings of the quantum map, we can consider a 
 smooth $\tau$-static wave function in the  eternal quantum mechanics, with width $\Gamma$ and centered around $\phi_0$. Its dual to the apocalyptic frame  has a narrowing width ${\widetilde \Gamma}(t) = \Gamma/\sqrt{\Omega(t)}$ as $t\rightarrow \pm \alpha\pi/2$, with its center migrating to the origin as $\tphi_0 (t) = \phi_0 /\sqrt{\Omega(t)}$, while at the same time the phase oscillates wildly. Therefore, the $\tPsi$ wave function is infinitely squeezed into the UV region (small $\tphi$)  as we approach the `apocalypse'. 

Conversely, starting with  $t$-static wave function with fixed width ${\widetilde \Gamma}$  and centered at $\tphi_0$  in the E-frame system, it corresponds to  an eternal wave function slipping into the deep IR (large $\phi$),  trailing the peak at $\phi_0 (\tau) = \tphi_0 \sqrt{\Omega(\tau)}$,  and widening at a rate of order $\Gamma(\tau) ={\widetilde \Gamma} \sqrt{\Omega(\tau)}$.

 To be more precise, let us focus on the operator map \opemap\ for the particular cases of interest.  
We shall adopt a terminology rooted in the behavior of the state in the apocalyptic frame (E-frame) as diagnosed by
the average values of polynomials in the canonical operators or natural observables such as the kinetic, potential and total energy. States with smooth E-frame behavior at $t=\pm t_\star$ can be continued beyond the `apocalyptic' times $\pm t_\star$ and will be termed {\it smooth} ($S$), while those with divergent matrix elements will be denoted as {\it singular}. Among the singular states, we shall refer to {\it crunches} ($C$) when the E-frame potential energy plummets to minus infinity:
$$
\lim_{|t| \to t_\star} \left\bra {\widetilde V} (\tphi\,) \right\ket_{\widetilde C} (t) = -\infty
\;.$$
Singular states with the opposite-sign divergence 
$$
\lim_{|t| \to t_\star} \left\bra {\widetilde V} (\tphi\,) \right\ket_{\widetilde B} (t) = +\infty
\;.$$
will be called {\it bubbles} ($B$) as analogs of `bubbles of nothing'.

The most interesting among singular states are those that look  {\it stationary} in the eternal frame, i.e. with a finite  $|\tau| \rightarrow \infty$ limit of $\bra V(\phi)\ket$, such as the `condensate states' considered in section 5. Any such state with a non-zero value of the eternal potential energy has an apocalyptic potential energy diverging  as $\Omega(t) \sim (t-t_\star)^{-\alpha}$, the hallmark of a singular state.

The anomalous transformation terms do affect the scaling of the kinetic energy:
\eqn\kna{
\left\bra \shalf \tpi^{\,2} \right\ket_{\tPsi} = \Omega(t)\left\bra \shalf \pi^2 \right\ket_{\Psi} -{1\over 4}   \pt_t \log\Omega \,\left\bra \{\phi, \pi\}\right\ket_{\Psi} + {1\over 8} \Omega^{-1} \left(\pt_t \log \Omega\right)^2 \left\bra \phi^2 \right\ket_{\Psi}
\;.}
For a condensate-type state in the eternal frame, the three terms in this equation scale as $(t-t_\star)^{-\alpha}$, $(t-t_\star)^{-1}$ and $(t-t_\star)^{\alpha-2}$ respectively. For either the EM or the EdS map, there is always a singular term for generic values of the eternal frame averages, confirming that the eternally stationary state is a singular state in the apocalyptic frame.    The anomalous terms (second and third on the right hand side of  \kna) are subdominant for the EM model ($\alpha=2$), but have the same scaling  as the first term in the EdS case ($\alpha=1$). \foot{This opens up the possibility that a certain dS-eternal state could be tuned to have finite kinetic energy in the E-frame, although other observables will still diverge in general.}

Starting with a smooth state in the E-frame, with finite and generic values of apocalyptic observables at $t=t_\star$,  the corresponding large-time behavior in the eternal frame follows from the inverse transformations. The 
potential energy vanishes as $\bra V(\phi)\ket \sim \Omega^{-1} \rightarrow 0$  as $|\tau| \rightarrow \infty$ for any value of $\alpha$ (recall this potential energy excludes the purely quadratic trapping term, to be considered below). We say that the state {\it dilutes} away in the eternal frame. 

The inverse of the \kna\ relation 
\eqn\knali{
\left\bra \shalf \pi^{\,2} \right\ket_{\Psi} = \Omega^{-1} \left\bra \shalf \tpi^2 \right\ket_{\tPsi} + {1\over 4}\Omega^{-1} \pt_t \log\Omega \,\left\bra \{\tphi, \tpi\}\right\ket_{\tPsi} + {1\over 8} \Omega^{-1} \left(\pt_t \log \Omega\right)^2 \left\bra \tphi^2 \right\ket_{\tPsi}
\;,}
implies a similar diluting behavior for the ordinary scaling term of the kinetic energy. The anomalous terms depending on derivatives of $\Omega$ have a potentially interesting behavior, since they scale as $(t-t_\star)^{\alpha-1}$ and $(t-t_\star)^{\alpha-2}$ respectively. Hence, for $\alpha=1$ (EdS map) we do get a divergent contribution to the kinetic energy in the $|\tau|\rightarrow \infty$ limit. 
We can understand this behavior by recalling that the EdS map relates the trapped CQM in the E-frame to
the anti-trapped (i.e. tachyonic) CQM in the dS-frame. This is characteristic of the $d=1$ case and implies
that smooth states look as falling down a harmonic cliff in the eternal frame. The potential energy coming from the trapping also diverges as
$$
-{1\over 8} \Omega(\tau) \,\bra \tphi^2 \ket_{\widetilde S} \longrightarrow -\infty\;,
$$
canceling the kinetic-energy infinity coming from the last term in \knali. Hence, the total energy does stay finite in the eternal runaway state. 

Even for $\alpha=2$, i.e. the EM map, the anomalous terms produce an interesting behavior, since the last one
yields an asymptotically constant value of the potential energy. This is to be understood as a state running away to large values of $\phi$, with asymptotically constant kinetic energy, i.e. a standard scattering state in the AFF quantum mechanical model. 

We conclude that the quantum complementarity map sends smooth E-frame states (which do not look `apocalyptic' in this frame) into states which run away towards large $\phi$ values in the eternal frame, with finite total energy
but diverging kinetic and potential components in the particular case of the EdS map.

Starting with a stationary state in the eternal frame, modeled as a `condensate' in the terminology of section 5,
the apocalyptic description carried by the $\tphi, \tpi$ operator algebra, sees it as a singular state. We say it is a crunch when the divergent potential energy is negative, and a bubble of nothing when it diverges to positive infinity.  
\subsec{Evanescent Crunches?}

\noindent

As previously indicated, the EdS map has peculiar properties related to the tachyonic character of the dS-frame Hamiltonian. 
Recall that the EdS map sends the standard trapped CQM (E-frame system) with Hamiltonian 
\eqn\et{
{\widetilde H} = \half \left(\tpi^2 + {\lambda \over \tphi^2} \right) + {1\over 8} \tphi^2 \;,
}
into the tachyonic CQM (dS-frame system) with Hamiltonian
\eqn\ea{
H = \half\left(\pi^2 + {\lambda \over \phi^2} \right) - {1\over 8} \phi^2 
}

The large-$\phi$ instability of the eternal-frame CQM implies that the standard negative deformations of the dS Hamiltonian 
$$
V_\Delta (\phi) = -{M^{1-\Delta} \over \phi^{2\Delta}}
$$
  with constant $M$ and $0<\Delta <1$, fail to induce a absolutely stable condensate at $\bra \phi \ket \sim M^{-1/2}$ with large $M$. In fact, the resulting states are only metastable, with a decay width of order $\Gamma \sim M \exp(-aM^{2/3})$ for some $\CO(1)$ constant $a$. Since the eternality of the condensate is related to the {\it crunchy} character of the state in the apocalyptic frame, it is interesting to inquire whether this metastability, inducing a finite life-time for the condensate, is capable of regularizing the crunch singularity. 
  
We can approximate the very-large $\tau$ wave-function of such states as
\eqn\vlt{
\Psi_{\rm meta} \approx e^{- \Gamma \tau/2} \Psi_{\rm cond} + \sqrt{1-e^{-\Gamma\tau}} \;\Psi_{\rm run}
}
where $\Psi_{\rm cond}$ is a normalized state which solves the Schr\"odinger equation in the large $M$ limit and represents the condensate
in the absence of the tachyonic instability, and $\Psi_{\rm run}$ is a state representing the runaway down the inverted harmonic potential 
after tunneling through the barrier. We can define this running state as the eternal dual from some generic finite-energy state in the E-frame system. 

Upon transforming this wave function to the E-frame system, the $\tPsi_{\rm cond}$ component has the characteristic crunchy behavior we mentioned above, whereas $\tPsi_{\rm run}$ is a smooth state in the apocalyptic frame. The amplitude of the crunchy component does vanish
in the $t\rightarrow \pi/2$ limit as
$$
e^{- \Gamma \tau/2} \sim |\,t_\star - t\,|^{\Gamma /2}
\;.$$
Since we have seen characteristic observables to diverge as inverse powers of $t-t_\star$, the contribution of $\tPsi_{\rm cond}$ to
expectation  values is of order
\eqn\exep{
|\,t-t_\star\,|^{\Gamma - b}
\;,}
where $b$ is some positive constant of  $\CO(1)$ whose detailed value depends on the particular observable being evaluated. In the semiclassical limit where this description of the tunneling is valid, 
 we have $\Gamma \sim M \exp(-a\,M^{2/3}) \ll 1$, so that $\Gamma \ll b$ and the quantum depletion of the condensate is not fast enough to turn off the crunchy behavior of the state. On the other hand, if the depletion rate should become of $\CO(1)$, the exponent
in 
\exep\ could change sign and the corresponding expectation value be smoothed out. We see that the potential for a quantum-mechanical smoothing of the crunch exists, by considering `condensates' with sufficiently fast decay rate. Formally, this situation can be engineered by tuning $M \ll 1$ in units of the background curvature. While the notion of `condensate' is not well defined in such a limit, it is worth mentioning that such states do exist in the dual AdS description (cf. the model discussed in the appendix of \refs\mald) and they have the same qualitative behavior as the more obvious crunch states described here. \foot{In fact, their bulk description is even simpler, since they can be analyzed as weak perturbations of AdS in supergravity.}

\newsec{Generalized Duality Between Eternity and Apocalypse}

\noindent

The description of conformal complementarity maps as quantum canonical transformations in CQM can be 
 {\it formally} extended to higher-dimensional field theories. Consider two conformally related $d$-dimensional Riemannian manifolds ${\bf X}$ and ${\bf\tX}$ with Weyl rescaling function $\Omega(x)$. Let us define a LG model on ${\bf X}$ with classical action 
\eqn\clac{
S_{\bf X}= -\int_{\bf X} \left[\shalf |\pt \phi |^2 +\shalf \,\xi_d \,\CR_{\bf X} \,\phi^2 +V( \phi)\right]\;.
}
The relevant perturbations
\eqn\mscal{
V(\phi) = \sum_i \varepsilon_i\,M_i^{\,d-\Delta_i}\,\phi^{{2\Delta_i \over d-2}}\;.
}
depend on  mass scales $M_i$. This model can be rewritten as a perturbed LG model   on ${\bf \tX}$ with action 
\eqn\clacd{
{ \widetilde S}_{\btX}= -\int_{\btX} \left[\shalf |\pt \tphi |^2 + \shalf\,\xi_d\, \CR_{\btX} \,\tphi^2 +
{\widetilde V} (\tphi\,) \right]\;,
}
plus some boundary terms. The
new potential reads  
$$
{\widetilde V} (\tphi\,) = \sum_i \varepsilon_i\,{\widetilde M}_i^{\,d-\Delta_i} \,\tphi^{\,{2\Delta_i \over d-2}}\;,
$$
in terms of  rescaled point-dependent mass scales ${\widetilde M}_i = \Omega \,M_i$ which now become `source-terms', and with the basic field redefinition $\tphi = \Omega^{d-2 \over 2} \,\phi$. 

The boundary terms are defined as 
$$
S_{\bf X}= { \widetilde S}_{\btX} -\Delta{\widetilde K} = {\widetilde S}_{\btX} +{d-2 \over 4} \int_{\pt \btX} \epsilon \ \,\left(\Omega^{-1} \, {\widetilde \nabla} \,\Omega\right)\,\tphi^{\,2}\;,
$$
where $\epsilon = -1$ for a space-like boundary component and $\epsilon = +1$ for a time-like boundary component and ${\widetilde \nabla}$ is the covariant derivative on ${\btX}$.

For the particular case of a Weyl rescaling function which is only dependent on time,  and a compact spatial section ${\bf K}$,  we can regard the ${\bf X}$ manifold as a cosmology 
$$
ds^2_{\bf X} = -d\tau^2 + \Omega(\tau)^2 \;d\ell^2_{\bf K}
\;,$$
while ${\btX}$ is a static cylinder with base ${\bf K}$:
$$
ds^2_{
\btX} = -dt^2 + d\ell^2_{\bf K}\;.
$$
If $\Omega$ maps a finite interval $t\in [-t_\star, t_\star]$ into the real line $\tau\in {\bf R}$ we can use the terminology that has become standard along this paper and regard ${\bf X}$ as the `eternal frame' and ${\btX}$ as the `apocalyptic frame'. Then, we only have
  space-like boundaries at $t=\pm t_\star$,
so that $\Delta {\widetilde K} = {\widetilde K} (t_\star) - {\widetilde K}(-t_\star)$, with 
\eqn\bttt{
{\widetilde K}(t)  = {d-2 \over 4}  \,\pt_t \log\,\Omega \int_{\bf K} \tphi^{\,2}\;.
}
The boundary term \bttt\ plays no significant role at the classical level, but does feature in the quantum treatment. The formal construction of the states and the canonical map parallels the previous formalism explained in section 5.2, except that Schr\"odinger-picture  wave-functionals replace wave-functions and canonical momenta are defined in terms of functional derivatives in the usual formal fashion, 
$
\pi(x) = -i \delta / \delta \phi(x)
$ and $\tpi(x) = -i\delta/\delta \tphi(x)$. This leads to analogous quantum maps generalizing \opemap:
\eqn\opemapd{\eqalign{
\left\langle F\left[\tphi\;;\; {\widetilde \pi}\,\right]\right\rangle_{\tPsi} &= \left\langle F\left[\Omega^{{d-2 \over 2}} \phi\;;\; \Omega^{2-d \over 2} \left(\pi +{d-2 \over 2}  \phi \,\Omega^{d-1}\pt_\tau \log \Omega\right) \right]\right\rangle_{\Psi}
\cr
\left\langle F\left[\phi\;;\;{ \pi}\,\right]\right\rangle_{\Psi} &= \left\langle F\left[\Omega^{2-d \over 2} \tphi\;; \;\Omega^{{d-2 \over 2}} \left({\widetilde \pi} -{d-2 \over 2} \tphi\, \pt_t \log \Omega\right) \right]\right\rangle_{\tPsi}
\;,}}
with an entirely similar interpretation as their $d=1$ counterparts. 

The Hamiltonian complementarity presented in \comh\ generalizes as well. The eternal-frame Hamiltonian can be written in the form
\eqn\efha{
H = \half \int_{\bf K} \Omega^{1-d}\,\pi^2  + \int_{\bf K} \Omega^{d-1} \left[\half \Omega^{-2} |\pt \phi |^2_{\bf K} + \half \xi_d \,\CR_{\bf X} \,\phi^2 + V(\phi) \right]\;,
}
and its apocalyptic counterpart:
\eqn\apha{
{\widetilde H} = \half \int_{\bf K} \tpi + \int_{\bf K} \left[ \half |\pt \tphi |^2_{\bf K} + \half \xi_d \,\CR_{ \btX} \,\tphi^{\,2}  + {\widetilde V} (\tphi\,)\right]
\;.}
Explicit calculation then shows that 
\eqn\cham{
\left[ H, {\widetilde H} \right] = 2i \,\CA(t)\, {\widetilde D}\;,
}
where ${\widetilde D} = \shalf \int_{\bf K} \{\tphi, \tpi\}$ and the anomalous  term generalizes to \foot{It is interesting that $\CA$ vanishes for $d=2$, a case that we have excluded consistently from our discussion in this paper. Precisely at $d=2$ we expect genuinely {\it anomalous} (quantum) contributions to \cham.}
\eqn\ant{
\CA(t) = {d-2 \over 4} \pt_t^2 \log \Omega + {(d-2)^2 \over 8} \left( \pt_t \log \Omega\right)^2 \;.
}
Although these relations are derived by simple canonical manipulations, we expect them to hold in all generality for general field theories, showing that the Hamiltonian complementarity induced by conformal mappings trivializes when acting on scale-invariant states, annihilated by the dilation operator. It is natural to interpret this result as underlying the fact that the global AdS vacuum is invariant under the action of the dilation operator, represented in the bulk as an isometry. On the other hand, for any state with an intrinsic scale, we expect a non-trivial quantum complementarity between the two Hamiltonian evolutions.

\newsec{Discussion}

\noindent

In this paper we have analyzed aspects of the general idea, going back at least to \refs{\banksdo, \insightfull}, that
complementarity maps can be realized as conformal transformations (or more general field-redefinitions) in
holographic models. A particularly simple example of this program was proposed in \refs\usd, in terms of condensate states on perturbed CFTs defined on dS space-time. A conformal map to the same CFT on the Einstein universe (E$_d$), but now perturbed by a time-dependent coupling, serves as the `infalling' frame in the sense of horizon complementarity. 

We have singled out the change of time variables, from an eternal history in dS$_d$, to a finite or `apocalyptic' one in E$_d$, as an `UV remnant' of the complementarity map, which can be studied using conventional Lagrangian methods. We have done so at the level of classical Landau--Ginzburg models of brane-like states, extending the analysis already presented in \refs\usd. A full quantum analysis is possible for the $d=1$ version, conformal quantum mechanics, which retains some qualitative properties akin to a AdS$_2$ dual, despite the fact that no actual reconstruction of bulk dynamics is available. In this case, the quantum map is a canonical transformation which rescales the canonical operator basis by a time-dependent factor. Once identified, the construction can be formally extended to higher dimensions. 

The existence of two operator algebras: $\{\CO\}$  for the  `exterior' observables and  $\{{\widetilde \CO}\}$ for  the `infalling' observables,  is similar to the case
of electric/magnetic duality or T-duality in string theory. In order to sharpen the analogy, we must enrich the models by allowing the coexistence of both `windings' and `momenta'. So far we have considered extremely simple states by way of example, such as either dS condensates or E-frame stationaries.
In order to realize the quantum complementarity map in a more physical fashion we must introduce an appropriate `measuring apparatus' for each operator algebra. 

Consider for instance a dS condensate state. Any physical system constructed from stationary states around the condensate ground state will only measure the eternal properties of the dS state. On the other hand, a physical system whose physical size is `comoving' with the Hubble expansion, will be able to `measure' a crunch in the $t$ time variable. It is tempting to us  to use our own universe as an example (in the limit $G_{\rm N} \rightarrow 0$ with fixed Hubble constant).
The Higgs condensate has fixed size in units of the Hubble constant, but a comoving `observer', anchored on the `realm of the nebulae',  will measure the 
$\{{\widetilde \CO}\}$ operator algebra. At large $\tau$-times such an observer is necessarily made of `neurons' separated by super-horizon distances, so that its workings appear completely non-local to an observer furnished with the $\{\CO\}$ operator algebra.  Conversely, in its own frame the $\{{\widetilde \CO}\}$ observer will see the
$\{\CO\}$ observer as a shrinking entity whose own Hamiltonian ramps up the eigenfrequencies to  produce the illusion of eternity in the face of an impending crunch.

The main limitation of these considerations is the absence of an actual reconstruction of operators with
approximate bulk locality, in the spirit of \refs{\banksdo, \bena, \lifsh, \polch,\papado}. In this sense, we have
strived to characterize horizon complementarity in the absence of an actual `horizon', using only deep UV data. The dichotomy between local and strongly non-local observables in the CFT should then become even more drastic when translated to reconstructed bulk operators.

\bigskip{\bf Acknowledgements:} 

The work J.L.F. Barbon  was partially supported by MEC and FEDER under a grants FPA2009-07908 and FPA2012-32828, the Spanish
Consolider-Ingenio 2010 Programme CPAN (CSD2007-00042), Comunidad Aut\'onoma de Madrid under grant HEPHACOS S2009/ESP-1473 and the 
spanish MINECO {\it Centro de Excelencia Severo Ochoa Program} under grant SEV-2012-0249. 
The work of E. Rabinovici  is partially supported by the
American-Israeli Bi-National Science Foundation,  the Israel Science Foundation Center
of Excellence and the I Core Program of the Planning and Budgeting Committee and The Israel 
Science Foundation "The Quantum Universe".

{\ninerm{
\listrefs
}}

\end